\documentclass[prd,
nofootinbib,
preprintnumbers,
twocolumn,
showpacs
]{revtex4}
\usepackage{amsmath}
\usepackage{amssymb}
\usepackage{epsfig}
\usepackage{graphicx}
\usepackage[normal]{subfigure}
\usepackage{rotating}

\newcommand{\capdef}{}
\newcommand{\mycaption}[2][\capdef]{\renewcommand{\capdef}{#2}
\caption[#1]{{\footnotesize #2}}}

\newcommand{\beq}{\begin{equation}}
\newcommand{\eeq}{\end{equation}}
\newcommand{\bdm}{\begin{displaymath}}
\newcommand{\edm}{\end{displaymath}}
\newcommand{\bea}{\begin{eqnarray}}
\newcommand{\eea}{\end{eqnarray}}
\newcommand{\nn}{\nonumber}

\def\eq#1{{Eq.~(\ref{#1})}}
\def\eqs#1#2{{Eqs.~(\ref{#1})--(\ref{#2})}}
\def\fig#1{{Fig.~\ref{#1}}}
\def\figs#1#2{{Figs.~\ref{#1}--\ref{#2}}}
\def\Table#1{{Table~\ref{#1}}}

\def\sect#1{{Sect.~\ref{#1}}}

\def\abs#1{\left| #1\right|}
\def\mod#1{\abs{#1}}

\def\Tr{\mbox{Tr}\,}

\def\chain#1#2{\mathrel{\mathop{\null\longrightarrow}\limits^{#1}_{#2}}}

\newcommand{\sepA}{\rule[-0.8cm]{0cm}{1.6cm}}
\newcommand{\sepB}{\rule[-1cm]{0cm}{2cm}}
\newcommand{\sepC}{\rule[-1.2cm]{0cm}{2.4cm}}

\begin{document}


\title{Intermediate mass scales in the non-supersymmetric $SO(10)$ grand unification:
a reappraisal}
\date{March 21, 2009}
\author{Stefano Bertolini}\email{bertolin@sissa.it}
\author{Luca Di Luzio}\email{diluzio@sissa.it}
\affiliation{INFN, Sezione di Trieste, and SISSA,
Via Beirut 4, I-34014 Trieste, Italy}
\author{Michal Malinsk\'{y}}\email{malinsky@kth.se}
\affiliation{Theoretical Particle Physics Group,
Department of Theoretical Physics,
Royal Technical Institute (KTH),
Roslagstullsbacken 21,
SE-106 91 Stockholm, Sweden.}
\begin{abstract}
The constraints of gauge unification on intermediate mass scales in
non-supersymmetric $SO(10)$ scenarios are systematically discussed.
With respect to the existing reference studies we
include the $U(1)$ gauge mixing renormalization at the one- and two-loop level,
and reassess the two-loop beta-coefficients.
We evaluate the effects of additional Higgs multiplets required at intermediate
stages by a realistic mass spectrum,
and update the discussion to the present day data.
On the basis of the obtained results, $SO(10)$ breaking patterns with up to two intermediate
mass scales are discussed for potential relevance and model predictivity.
\vspace*{0ex}
\end{abstract}
\maketitle
\renewcommand{\thefootnote}{\alph{footnote}}
\renewcommand{\thefootnote}{\fnsymbol{footnote}}
\renewcommand{\thefootnote}{\it\alph{footnote}}
\renewcommand{\thefootnote}{\arabic{footnote}}
\setcounter{footnote}{0}

\section{Introduction}
\label{sec:intro}

Understanding theoretically
the patterns of masses and mixings of ordinary fermions is one of the
long aimed goals in particle physics. Of the 56 parameters in the
Standard Model (SM) Yukawa sector (including Majorana neutrinos) only
22 can be measured at low energy and just 17 have been determined from the experiment.
Grand Unified Theories (GUTs),
by enforcing stringent
relations among the different particle sectors and by reducing the degeneracy
in the parameter space,
do provide a powerful tool for
addressing the multiplicity of matter states and the detailed structure
of the Yukawa sector.

Appealing candidates for realistic GUTs are models based on the $SO(10)$
gauge group~\cite{GUT}.
All the known SM fermions plus three right-handed neutrinos fit into three
copies of the 16-dimensional spinorial representation of $SO(10)$, thus providing a rationale for the SM hypercharge structure. The
model also provides a natural explanation for the sub-eV light neutrino masses via the seesaw mechanism~\cite{seesawI,seesawII}.

The purpose of this paper is to review the constraints enforced by gauge unification
on the intermediate mass scales in the non-supersymmetric $SO(10)$ GUTs, a needed
preliminary step for assessing the structure of
the multitude of the different breaking patterns
before entering the details of a specific model.
Eventually, our goal is to envisage and examine scenarios potentially
relevant for the understanding of the low energy matter spectrum.
In particular those setups that, albeit
non-supersymmetric, may exhibit a predictivity comparable to that of the minimal supersymmetric $SO(10)$, scrutinized at length in the last few years~\cite{MSSO10}.

The most recent discussion of fermion masses and mixings in non-supersymmetric $SO(10)$ GUTs was given in Ref.~\cite{Bajc:2005zf}. The authors focussed only on renormalizable models
(i.e. without the spinorial $\overline{16}_H$ in the Higgs sector)
with combinations of $10_H$ and $\overline{126}_H$ or $120_H$ driving the Yukawa interactions.
Particular attention is paid to the leptonic sector and the mechanism of generation
of neutrino masses via see-saw.

The constraints imposed by the absolute neutrino
mass scale on the position of the $B-L$ threshold, together with the proton decay bound on the unification scale $M_{U}$,  provide a
discriminating tool among the many $SO(10)$ scenarios and the corresponding breaking patterns.
These were studied at length in the eighties and early nineties, and detailed surveys
of two- and three-step $SO(10)$ breaking chains
(one and two intermediate thresholds respectively)
are found in Refs.~\cite{Gipson:1984aj,Chang:1984qr,Deshpande:1992au,Deshpande:1992em}.

We perform a systematic survey of $SO(10)$ unification with two intermediate stages.
In addition to updating the analysis to present day data,
this reappraisal is motivated by
(a) the absence of $U(1)$ mixing in previous studies, both at one- and two-loops in the gauge coupling renormalization,
(b) the need for additional Higgs multiplets at some intermediate stages,
and
(c) a reassessment of the two-loop beta coefficients reported in the literature.

The outcome of our study is the emergence of sizeably different features in some of the breaking patterns as compared to the existing results.
This allows us to rescue previously excluded scenarios.
All that before considering the effects of
threshold corrections~\cite{Dixit:1989ff,Mohapatra:1992dx,Lavoura:1993su},
that are unambiguously assessed only when the details of a specific model are worked out.

It is remarkable that the chains corresponding to the minimal $SO(10)$ setup with the smallest Higgs representations ($10_H$, $45_H$ and $\overline{16}_H$, or $\overline{126}_H$
in the renormalizable case) and the smallest number of parameters in the
Higgs potential, are still viable. The complexity of this non-supersymmetric scenario is comparable to that of the minimal supersymmetric $SO(10)$ model, what makes it worth of detailed consideration.

In \sect{sec:chains} we set the framework of the analysis.
\sect{sec:2Lrge} provides a collection of the tools needed for a two-loop study of grand unification.
The results of the numerical study are reported and scrutinized in \sect{sec:results}.
Perspectives for further progress are discussed in \sect{sec:outlook}.
Finally, the relevant one- and two-loop $\beta$-coefficients are detailed
in Appendix \ref{app:2Lbeta}.

\section{Three-step $SO(10)$ breaking chains}
\label{sec:chains}

The relevant $SO(10)\to G2\to G1\to SM$ symmetry breaking chains with two intermediate
gauge groups $G2$ and $G1$ are listed in Table~\ref{tab:chains}. Effective two-step chains are obtained
by identifying two of the high-energy scales, paying attention to the possible
deviations from minimality of the scalar content in the remaining intermediate stage
(this we shall discuss in \sect{sec:2loopgauge}).

For the purpose of comparison we follow closely the notation of ref.~\cite{Deshpande:1992em},
where $P$ denotes the unbroken D-parity~\cite{Chang:1983fu}.
For each step the Higgs representation responsible for the
breaking is given.

The breakdown of the lower intermediate symmetry $G1$ to the SM gauge group is driven either by the
$16$- or $126$-dimensional Higgs multiplets  $\overline{16}_{H}$ or  $\overline{126}_{H}$.
An important feature of the scenarios with $\overline{126}_{H}$
is the fact that in such a case a potentially realistic $SO(10)$ Yukawa
sector can be constructed already at the renormalizable level.
Together with $10_{H}$ all the
effective Dirac Yukawa couplings as well as the Majorana mass matrices at the SM level
emerge from the contractions of the matter bilinears $16_{F}16_{F}$ with $\overline{126}_H$
or with $\overline{16}_H \overline{16}_H/\Lambda$, where $\Lambda$
denotes the scale (above $M_{U}$)
at which the effective dimension five Yukawa couplings arise.

\begin{table}
\begin{tabular}{lllll}
\hline
Chain  &  & G2 & & G1
\\
\hline
{\rm I:}  &
           $ \chain{}{210}  $ & $ \{2_L 2_R 4_C\} $
         & $ \chain{}{45}  $ & $ \{2_L 2_R 1_X 3_c\} $
\\
{\rm II:} &
           $ \chain{}{54}  $ & $ \{2_L 2_R 4_C P\} $
         & $ \chain{}{210}   $ & $ \{2_L 2_R 1_X 3_c P\} $
\\
{\rm III:}  &
           $ \chain{}{ 54 }  $ & $ \{2_L 2_R 4_C P\} $
         & $ \chain{}{ 45 }  $ & $ \{2_L 2_R 1_X 3_c\} $
\\
{\rm IV:}  &
           $ \chain{}{ 210 }  $ & $ \{2_L 2_R 1_X 3_c P\} $
         & $ \chain{}{ 45 }  $ & $ \{2_L 2_R 1_X 3_c\} $
\\
{\rm V:}  &
           $ \chain{}{ 210 }  $ & $ \{2_L 2_R 4_C\} $
         & $ \chain{}{45}  $ & $ \{2_L 1_R 4_C\} $
\\
{\rm VI:}  &
           $ \chain{}{ 54 }  $ & $ \{2_L 2_R 4_C P\} $
         & $ \chain{}{ 45 }  $ & $ \{2_L 1_R 4_C\} $
\\
{\rm VII:}  &
           $ \chain{}{ 54 }  $ & $ \{2_L 2_R  4_C P\} $
         & $ \chain{}{ 210 }  $ & $ \{2_L 2_R 4_C\} $
\\
{\rm VIII:}  &
           $ \chain{}{ 45 }  $ & $ \{2_L 2_R 1_X 3_c\} $
         & $ \chain{}{ 45 }   $ & $ \{2_L 1_R 1_X 3_c\} $
\\
{\rm IX:}  &
           $ \chain{}{ 210 }  $ & $ \{2_L 2_R 1_X 3_c P\} $
         & $ \chain{}{ 45 }  $ & $ \{2_L 1_R 1_X 3_c\} $
\\
{\rm X:}  &
           $ \chain{}{ 210 }  $ & $ \{2_L 2_R 4_C\} $
         & $ \chain{}{  210 } $ & $ \{2_L 1_R 1_X  3_c\} $
\\
{\rm XI:}  &
           $ \chain{}{ 54 }  $ & $ \{2_L 2_R 4_C P\} $
         & $ \chain{}{ 210 }  $ & $ \{2_L 1_R 1_X 3_c\} $
\\
{\rm XII:}  &
           $ \chain{}{ 45 }  $ & $ \{2_L 1_R 4_C\} $
         & $ \chain{}{ 45 }  $ & $ \{2_L 1_R 1_X 3_c\} $
\\[1ex]
\hline
\end{tabular}
\mycaption{Relevant $SO(10)$ symmetry breaking chains via two intermediate
gauge groups G1 and G2. For each step the representation of the Higgs multiplet
(in $SO(10)$ notation) responsible for the breaking is given.
The breaking to the SM group $1_Y 2_L 3_c$ is obtained via a $16$ or $126$ Higgs representation. The naming and ordering of the gauge groups follows the notation of ref.~\cite{Deshpande:1992em}.}
\label{tab:chains}
\end{table}

The Higgs transforming as $10$ under $SO(10)$ may carry in general extra quantum numbers
of a complex representation of some additional symmetry (a discussion on the implementation
of a Peccei-Quinn $U(1)_{PQ}$ symmetry in this scenario is given in Ref.~\cite{Bajc:2005zf}).
In this case it is sufficient to consider only two complex symmetric matrices $Y_{10}$ and $Y_{126}$ at the renormalizable $SO(10)$ level, namely
\begin{equation}
\label{minimalSO10Yukawasector}
16_{F}(Y_{10} 10_{H}+Y_{126} \overline{126}_{H})16_{F} \ ,
\end{equation}
that govern all the effective Yukawa couplings at lower energies.
Such scenarios are rather constrained and hence their detailed numerical studies are well motivated .

D-parity is a discrete symmetry acting as charge conjugation in a left-right
symmetric context~\cite{Chang:1983fu}, and as that it plays the role of a left-right symmetry (it enforces for instance equal left and right gauge
couplings).
$SO(10)$ invariance then implies exact D-parity (because D belongs
to the $SO(10)$ Lie algebra). D-parity may be spontaneously broken by D-odd Pati-Salam (PS)
singlets contained in 210 or 45 Higgs representations. Its breaking can therefore be decoupled
from the $SU(2)_R$ breaking, allowing for different left and right gauge couplings.

\begin{table*}[ht]
\begin{tabular}{cccccc}
\hline
& \multicolumn{4}{c}{Surviving Higgs multiplets in $SO(10)$ subgroups} & \\
\cline{2-5}
\multicolumn{1}{c}{$SO(10)$}&
\multicolumn{1}{c}{$\{2_L 1_R4_C\}$} &
\multicolumn{1}{c}{$ \{2_L 2_R 4_C\}$} &
\multicolumn{1}{c}{$\{2_L 2_R 1_X 3_c\}$} &
\multicolumn{1}{c}{$\{2_L 1_R 1_X 3_c\}$} &
\multicolumn{1}{c}{Notation}\\
\hline
10 & $(2, {+\frac{1}{2}}, 1)$ & $(2,2,1)$ & $(2,2,0,1)$ &
$(2,{+\frac{1}{2}},0,1)$ &
$\phi^{10}$\\
$\overline{16}$ & $(1, +{\frac{1}{2}},{4})$ & $(1,2,{4})$ & $(1,2,{-\frac{1}{2}},1)$ &
$(1,+{\frac{1}{2}},{-\frac{1}{2}},1)$ &
$\delta^{16}_R$\\
$\overline{16}$ &  & $(2,1,\overline{4})$ & $(2,1,+{\frac{1}{2}},1)$ &  &
$\delta^{16}_L$\\
$\overline{126}$ & $(2,+\frac{1}{2},15)$  & $(2,2,15)$  & $(2,2,0,1)$ &
$(2,{+\frac{1}{2}},0,1)$ &
$\phi^{126}$\\
$\overline{126}$ &$(1,1,10)$  & $(1,3,10)$ & $(1,3,-1,1)$ & $(1,1,-1,1)$ &
$\Delta^{126}_R$\\
$\overline{126}$ &  & $(3,1,\overline{10})$ & $(3,1,1,1)$ & &
$\Delta^{126}_L$\\
45 & $(1,0,15)$ & $(1,1,15)$ &  & &
$\Lambda^{45}$\\
210 &  & $(1,1,15)$ &  & &
$\Lambda^{210}$\\
45 &  & $(1,3,1)$ &$(1,3,0,1)$  & &
$\Sigma^{45}_R$\\
45 &  & $(3,1,1)$ &$(3,1,0,1)$  & &
$\Sigma^{45}_L$\\
210 &  & $(1,3,15)$ &  & &
$\sigma^{210}_R$\\
210 &  & $(3,1,15)$ &  & &
$\sigma^{210}_L$\\
\hline
\end{tabular}
\mycaption{Scalar multiplets contributing to the running of the
gauge couplings for a given $SO(10)$ subgroup according to minimal fine tuning.
The survival of $\phi^{126}$ (not required by minimality) is needed by
a realistic leptonic mass spectrum, as discussed in the text
(in the $2_L 2_R 1_X 3_c$ and $2_L 1_R 1_X 3_c$ stages only one linear combination
of $\phi^{10}$ and $\phi^{126}$ remains).
The $U(1)_X$ charge is given, up to a factor $\sqrt{3/2}$, by $(B-L)/2$
(the latter is reported in the table). For the naming of the
Higgs multiplets we follow the notation of Ref.~\cite{Deshpande:1992em}
with the addition of $\phi^{126}$.
When the D-parity (P) is unbroken the particle content must be left-right symmetric.
D-parity may be broken via P-odd Pati-Salam singlets in $45_H$ or $210_H$.}
\label{tab:submultiplets}
\end{table*}

The possibility of decoupling the D-parity breaking from the scale of right-handed interactions is a cosmologically
relevant issue. On the one hand baryon asymmetry cannot arise in a left-right symmetric
($g_L=g_R$) universe~\cite{Kuzmin:1980yp}.
On the other hand, the spontaneous breaking of a discrete symmetry, such as D-parity,
creates domain walls that, if massive enough (i.e. for intermediate mass scales) do not disappear, overclosing the universe~\cite{Kibble:1982dd}.
These potential problems may be overcome either
by confining D-parity at the GUT scale or by invoking inflation. The latter solution implies
that domain walls are formed above the reheating temperature, enforcing a lower bound
on the D-parity breaking scale of $10^{12}$ GeV. Realistic $SO(10)$ breaking
patterns must therefore include this constraint.

\subsection{The extended survival hypothesis}
\label{sec:ESH}

Throughout all three stages of running we assume that the scalar spectrum obeys the so
called extended survival hypothesis (ESH)~\cite{del Aguila:1980at}
which requires that {\it at every stage of the symmetry breaking chain only those scalars
are present that develop a vacuum expectation value (VEV)
at the current or the subsequent levels of the spontaneous symmetry breaking}.
ESH is equivalent to the requirement of the minimal number of
fine-tunings to be imposed onto the scalar potential~\cite{Mohapatra:1982aq}
so that all the symmetry breaking steps are performed at the desired scales.

On the technical side one should identify all the Higgs multiplets needed by the
breaking pattern under consideration and keep them
according to the gauge symmetry down to the scale of their VEVs.
This typically pulls down a large number of scalars in scenarios where
$\overline{126}_{H}$ provides the $B-L$ breakdown.

On the other hand, one must take into account that the role of $\overline{126}_{H}$ is twofold: in addition to triggering the $G1$ breaking it plays a relevant role
in the Yukawa sector (\eq{minimalSO10Yukawasector})
where it provides the necessary breaking of
the down quark - charged lepton mass degeneracy.
For this to work one needs a reasonably large admixture of the $\overline{126}_{H}$
component in the effective electroweak doublets. Since $(2,2,1)_{10}$ can mix
with $(2,2,15)_{\overline{126}}$ only below the Pati-Salam breaking scale, both fields
must be present at the Pati-Salam level (otherwise the scalar doublet mass matrix does not provide large enough components of both these multiplets in the light Higgs fields).

Note that the same argument applies also to the $2_{L}1_{R}4_{C}$ intermediate stage
when one must retain the doublet component of $\overline{126}_{H}$, namely $(2,+\frac{1}{2},15)_{\overline{126}}$,
in order for it to eventually admix with  $(2,+\frac{1}{2},1)_{10}$
in the light Higgs sector.
On the other hand, at the $2_L 2_R 1_X 3_c$ and $2_L 1_R 1_X 3_c$ stages,
the (minimal) survival of only one combination of the $\phi^{10}$
and $\phi^{126}$ scalar doublets (see Table \ref{tab:submultiplets}) is compatible with the Yukawa sector constraints because the degeneracy between the quark and lepton spectra has already been smeared-out by the Pati-Salam breakdown.

In summary, potentially realistic renormalizable Yukawa textures in settings with
well-separated $SO(10)$ and Pati-Salam breaking scales
call for an additional fine tuning in the Higgs sector.
In the scenarios with $\overline{126}_{H}$,
the $10_{H}$ bidoublet $(2,2,1)_{10}$,
included in Refs~\cite{Gipson:1984aj,Chang:1984qr,Deshpande:1992au,Deshpande:1992em},
must be paired at the $2_{L}2_{R}4_{C}$ scale with an
extra $(2,2,15)_{\overline{126}}$ scalar bidoublet
(or $(2,+\frac{1}{2},1)_{10}$ with $(2,+\frac{1}{2},15)_{\overline{126}}$ at the $2_{L}1_{R}4_{C}$ stage).
This can affect the running of the gauge couplings in chains I, II, III, V, VI, VII, X, XI
and XII.

For the sake of comparison with previous
studies~\cite{Gipson:1984aj,Chang:1984qr,Deshpande:1992au,Deshpande:1992em}
we shall not include the $\phi^{126}$ multiplets in the first part of the analysis.
Rather, we shall comment on their relevance for gauge unification in \sect{sec:extrahiggs}.

\section{Two-loop gauge renormalization group equations}
\label{sec:2Lrge}

In this section we report, in order to fix a consistent notation, the two-loop renormalization group equations
(RGEs) for the gauge couplings.
We consider a gauge group of the form 
$ U(1)_{1} \otimes ... \otimes U(1)_{N}\otimes G_1\otimes ... \otimes G_{N'}$, where $G_i$ are simple groups.

\subsection{The non-abelian sector}
\label{sec:2Lnonabelian}

Let us focus first on the non-abelian sector  corresponding to $G_1\otimes ... \otimes G_{N'}$  and defer the full treatment of the effects due to the extra $U(1)$ factors to section \ref{sec:U1mix}.
Defining $t=\log (\mu/\mu_0)$ we write
\beq
\frac{dg_p}{dt} = g_p\ \beta_p
\label{rge}
\eeq
where $p=1,...,N'$ is the gauge group label. Neglecting for the time being the abelian components,
the $\beta$-functions for the $G_1\otimes ...\otimes G_{N'}$ gauge couplings  read at two-loop level~\cite{Jones:1974mm,Caswell:1974gg,Jones:1981we,Machacek:1983-85}:
\bea
\beta_p &=& \frac{g_p^2}{(4\pi)^2} \left\{
    - \frac{11}{3} C_2(G_p) + \frac{4}{3}\kappa S_2(F_p) + \frac{1}{3} \eta S_2(S_p) \right. \nn \\
&-&   \frac{2\kappa}{(4\pi)^2} Y_4(F_p)
    + \frac{g_p^2}{(4\pi)^2} \left[- \frac{34}{3} \left( C_2(G_p) \right)^2  \right.     \nn \\
&+&                \left( 4 C_2(F_p) + \frac{20}{3} C_2(G_p) \right) \kappa S_2(F_p)     \label{Gp2loops} \\
&+&         \left. \left( 4 C_2(S_p) + \frac{2}{3} C_2(G_p) \right) \eta S_2(S_p) \right] \nn \\
&+&         \left.\frac{g_q^2}{(4\pi)^2} 4 \Big[  \kappa C_2(F_q) S_2(F_p)
                         + \eta C_2(S_q) S_2(S_p)  \Big] \!\right\},                \nn
\eea
where $\kappa=1,\frac{1}{2}$ for Dirac and Weyl fermions respectively.
Correspondingly, $\eta=1, \frac{1}{2}$ for complex and real scalar fields. The sum over $q\neq p$ corresponding to contributions to $\beta_{p}$ from the other gauge sectors labelled by $q$ is understood.
Given a fermion $F$ or a scalar $S$ field that transforms according to the
representation $R=R_1\otimes ... \otimes R_{N'}$, where $R_p$ is an irreducible
representation of the group $G_p$ of dimension $d(R_p)$, the factor $S_2(R_p)$ is defined by
\beq
S_2(R_p) \equiv T(R_p)\frac{d(R)}{d(R_p)}\,,
\label{S2}
\eeq
where $T(R_p)$ is the Dynkin index of the representation $R_p$. The corresponding Casimir
eigenvalue is then given by
\beq
C_2(R_p) d(R_p) = T(R_p) d(G_p)\,,
\label{C2}
\eeq
where $d(G)$ is the dimension of the group.
In \eq{Gp2loops} the first row represents the one-loop contribution while the other
terms stand for the two-loop corrections, including that induced by Yukawa interactions.
The latter is accounted for in terms of a factor
\beq
Y_4(F_p) = \frac{1}{d(G_p)} \Tr \left[ C_2(F_p) YY^\dagger \right],
\label{Y4}
\eeq
where the ``general'' Yukawa coupling
\beq
Y^{abc}\ \overline{\psi}_a \psi_b\ h_c \ + \ h.c.
\label{Yukawa}
\eeq
includes family as well as group indices.
The coupling in \eq{Yukawa} is written in terms of four-component Weyl spinors $\psi_{a,b}$
and a scalar field $h_c$ (be complex or real). The trace includes the sum over all relevant fermion and
scalar fields.

\vspace*{3ex}
\subsection{The abelian couplings and $U(1)$ mixing}
\label{sec:U1mix}

In order to include the abelian contributions to \eq{Gp2loops} at two loops and the
one- and two-loop effects of $U(1)$ mixing~\cite{Holdom:1985ag}, let us write the most general
interaction of $N$ abelian gauge bosons $A^\mu_b$ and a set of
Weyl fermions $\psi_f$ as
\beq
\overline{\psi_f} \gamma_\mu Q^r_f\psi_f g_{rb} A^\mu_b\,.
\label{mincoupl}
\eeq
The gauge coupling constants $g_{rb}$, $r,b=1,...,N$, couple $ A^\mu_b$ to the fermionic
current $J^r_\mu = \overline{\psi}_f \gamma_\mu Q^r_f\psi_f$. The $N\times N$ gauge coupling
matrix $g_{rb}$ can be diagonalized by two independent rotations: one acting on the $U(1)$ charges
$Q^r_f$ and the other on the gauge boson fields $A^\mu_b$.
For a given choice of the charges,
$g_{rb}$ can be set in a triangular form ($g_{rb}=0$ for $r>b$) by the gauge boson rotation. The resulting
$N(N+1)/2$ entries are observable couplings.

Since $F^a_{\mu\nu}$ in the abelian case is itself gauge invariant, the most general kinetic part of the lagrangian reads at the renormalizable level
\beq
-\frac{1}{4} F^a_{\mu\nu}  F^{a\mu\nu}  - \frac{1}{4} \xi_{ab} F^a_{\mu\nu}  F^{b\mu\nu}\,,
\label{FmunuFmunu}
\eeq
where $a\neq b$ and $\mod{\xi_{ab}} < 1$.
A non-orthogonal rotation of the fields $A^\mu_a$ may be performed to set the gauge kinetic term
in a canonical diagonal form. Any further orthogonal rotation of the gauge fields will preserve
this form. Then, the renormalization prescription may be conveniently chosen to
maintain at each scale the kinetic terms canonical and diagonal on-shell while renormalizing
accordingly the gauge coupling matrix $g_{rb}$~\footnote{Alternatively
one may work with off-diagonal kinetic terms while keeping the
gauge interactions diagonal~\cite{Luo:2002iq}.}.
Thus, even if at one scale $g_{rb}$ is diagonal, in general non-zero off-diagonal
entries are generated by renormalization effects. One shows~\cite{delAguila:1988jz}
that in the case
the abelian gauge couplings are at a given scale diagonal {\em and} equal (i.e. there is a $U(1)$ unification),
there may exist a (scale independent) gauge field basis such that the abelian interactions
remain to all orders diagonal along the RGE trajectory~\footnote{Vanishing of the commutator of the $\beta$-functions and their derivatives is needed~\cite{delAguila:1995rb}.}.

In general, the renormalization of the abelian part of the gauge interactions
is determined by
\beq
\frac{dg_{rb}}{dt} = g_{ra} \beta_{ab}\,,
\label{U1rge}
\eeq
where, as a consequence of gauge invariance,
\beq
\beta_{ab} = \frac{d}{dt}\left(\log Z^{1/2}_3 \right)_{ab}\,.
\label{U1beta}
\eeq
with $Z_3$ denoting the gauge-boson wave-function renormalization matrix.
In order to further simplify the notation
it is convenient to introduce the ``reduced'' couplings~\cite{delAguila:1988jz}
\beq
g_{kb} \equiv Q^r_k g_{rb}\,,
\label{redcoupl}
\eeq
that evolve according to
\beq
\frac{dg_{kb}}{dt} = g_{ka} \beta_{ab}\,.
\label{U1rgered}
\eeq
The index $k$ labels the fields (fermions and scalars) that carry $U(1)$ charges.

In terms of the reduced couplings the $\beta$-function that governs the $U(1)$
running up to two loops is given
by~\cite{Jones:1974mm,Caswell:1974gg,Jones:1981we}
\bea
\beta_{ab} &=& \frac{1}{(4\pi)^2} \left\{
      \frac{4}{3}\kappa\ g_{fa}g_{fb} + \frac{1}{3}\eta\ g_{sa}g_{sb}     \right.       \nn \\
&-&   \frac{2\kappa}{(4\pi)^2}   \Tr \left[g_{fa}g_{fb}\ YY^\dagger \right]\nn \\
&+&   \frac{4}{(4\pi)^2} \Big[
    \kappa \left( g_{fa}g_{fb}g_{fc}^2 + g_{fa}g_{fb} g_q^2 C_2(F_q) \right)        \nn \\
&+&   \left. \eta \left( g_{sa}g_{sb}g_{sc}^2 + g_{sa}g_{sb} g_q^2 C_2(S_q) \right) \Big] \right\}\,,
\label{bU12loops}
\eea
where repeated indices are summed over, labelling fermions ($f$), scalars ($s$)
and $U(1)$ gauge groups ($c$). The terms proportional to the quadratic Casimir $C_2(R_p)$ represent the two-loop
contributions of the non abelian components $G_q$ of the gauge group to the $U(1)$ gauge coupling
renormalization.

Correspondingly, using the notation of \eq{redcoupl}, an additional two-loop term that represents the renormalization of the non abelian gauge couplings
induced at two loops by the $U(1)$ gauge fields is to be added
to \eq{Gp2loops}, namely
\bea
\Delta \beta_p &=& \frac{g_p^2}{(4\pi)^4}
                4 \Big[ \kappa\ g_{fc}^2 S_2(F_p) +  \eta\ g_{sc}^2 S_2(S_p)\Big]\,.\;\;\;\;
\label{bpU12loops}
\eea
In \eqs{bU12loops}{bpU12loops}, we use the abbreviation $f\equiv F_p$ and $s\equiv S_p$ and, as before, $\kappa=1,\frac{1}{2}$ for Dirac and Weyl fermions, while $\eta=1,\frac{1}{2}$ for complex and real scalar fields respectively.

\subsection{Some notation}
\label{sec:2Lnotation}

When at most one $U(1)$ factor is present, and neglecting the Yukawa contributions,
the two-loop RGEs can be conveniently written as
\beq
\frac{d\alpha_i^{-1}}{dt} = - \frac{a_i}{2\pi} - \frac{b_{ij}}{8\pi^2}\alpha_j\,,
\label{alpharge}
\eeq
where $\alpha_i=g_i^2/4\pi$. The $\beta$-coefficients $a_i$ and $b_{ij}$ for the
relevant $SO(10)$ chains are given  in Appendix \ref{app:2Lbeta}.

Substituting the one-loop solution for $\alpha_j$ into the right-hand side of \eq{alpharge}
one obtains
\beq
\alpha_i^{-1}(t) - \alpha_i^{-1}(0) = - \frac{a_i}{2\pi}\ t + \frac{\tilde b_{ij}}{4\pi}
                                     \log\left(1- \omega_j t\right)\,,
\label{alpha2loops}
\eeq
where $\omega_j=a_j \alpha_j(0)/(2\pi)$ and $\tilde b_{ij} = {b_{ij}}/{a_j}$ .
The analytic solution in (\ref{alpha2loops}) holds at two loops
(for $\omega_j t < 1$) up to higher order
effects. A sample of the rescaled $\beta$-coefficients $\tilde b_{ij}$ is given,
for the purpose of comparison with previous results,
in Appendix~\ref{app:2Lbeta}.

We shall conveniently write the $\beta$-function in \eq{bU12loops},
that governs the abelian mixing, as
\bea
\beta_{ab} &=& \frac{1}{(4\pi)^2}\ g_{sa}\ \gamma_{sr}\ g_{rb}\,,
\label{bU1gamma}
\eea
where $\gamma_{sr}$ include both one- and two-loop contributions.
Analogously, the non-abelian beta function in \eq{Gp2loops},
including the $U(1)$ contribution in \eq{bpU12loops},
is conveniently written as
\bea
\beta_{p} &=& \frac{g_{p}^2}{(4\pi)^2}\ \gamma_{p}\,.
\label{bU1gammanonabel}
\eea
The $\gamma_{p}$ functions for the $SO(10)$ breaking chains considered in this work are reported
in Appendix~\ref{app:U1mix}.

Finally, the Yukawa term in \eq{Y4}, and correspondingly in \eq{bU12loops}, can be written as
\beq
Y_4(F_p) =  y_{pk}\Tr \left(Y_kY_k^\dagger \right)\,,
\label{Ycpk}
\eeq
where $Y_{k}$ are the ``standard'' $3\times 3$ Yukawa matrices in the family space labelled by the flavour index $k$. The trace is taken over family indices and $k$ is summed over the different Yukawa
terms present at each stage of $SO(10)$ breaking. The coefficients $y_{pk}$ are given
explicitly in Appendix \ref{app:Yukawa}

\subsection{One-loop matching}
\label{sec:1Lmatching}

The matching conditions between effective theories in the framework of
dimensional regularization have been derived in \cite{Weinberg:1980wa,Hall:1980kf}.
Let us consider first a simple gauge group $G$ spontaneously broken into subgroups $G_p$.
Neglecting terms involving logarithms of mass ratios which are expected
to be subleading (massive states clustered near the threshold\footnote{An early discussion of thresholds effects in $SO(10)$ GUT is found in \cite{Dixit:1989ff}.}),
the one-loop matching for the gauge couplings can be written as
\beq
\alpha_{p}^{-1} - \frac{C_2(G_p)}{12\pi} = \alpha_G^{-1} - \frac{C_2(G)}{12\pi}\,.
\label{2LmatchGUT}
\eeq
Let us turn to the case when several non-abelian simple groups $G_p$
(and at most one $U(1)_X$)
spontaneously break whilst preserving a $U(1)_Y$ charge. The conserved $U(1)$ generator $T_Y$
can be written in terms of the relevant generators of the various Cartan subalgebras (and of the consistently
normalized $T_X$)
as
\beq
T_Y = p_i T_i\,,
\label{oneU1}
\eeq
where $\sum p_i^2 = 1$, and $i$ runs over the relevant $p$ (and $X$) indices.
The matching condition is then given by
\beq
\alpha_Y^{-1} = \sum_i p_i^2 \left(\alpha_i^{-1} - \frac{C_2(G_i)}{12\pi}\right)\,,
\label{2LmatchNOmix}
\eeq
where for $i=X$, if present, $C_2=0$.

Consider now the breaking of $N$ copies of $U(1)$ gauge factors to a subset of $M$ elements $U(1)$ (with $M<N$).
Denoting by $T_n$ ($n=1,...,N$) and by $\widetilde T_m$ ($m=1,...,M$)
their properly normalized generators we have
\beq
\widetilde T_m = P_{mn} T_n
\label{U1charges}
\eeq
with the orthogonality condition $P_{mn}P_{m'n} = \delta_{mm'}$. Let us denote
by $g_{na}$ ($n,a=1,...,N$) and by $\widetilde g_{mb}$ ($m,b=1,...,M$) the matrices
of abelian gauge couplings above and below the breaking scale respectively.
By writing the abelian gauge boson mass matrix in the broken vacuum and by identifying
the massless states, we find the following matching condition
\beq
(\widetilde g \widetilde g^T )^{-1} = P \left(g g^T \right)^{-1} P^T\,.
\label{2LU1match}
\eeq
Notice that \eq{2LU1match} depends on the chosen basis for the $U(1)$ charges (via $P$)
but it is invariant under orthogonal rotations of the gauge boson fields ($gO^TOg^T=gg^T$).
The massless gauge bosons $\widetilde A_m^\mu$ are given in terms of $A_n^\mu$ by
\beq
\widetilde A_m^\mu = \left[ \widetilde g^T P \left(g^{-1}\right)^T \right]_{mn} A_n^\mu\,,
\label{masslessA}
\eeq
where $m=1,...,M$ and $n=1,...,N$.

The general case of a gauge group $ U(1)_{1} \otimes ... \otimes U(1)_{N}\otimes G_1\otimes ... \otimes G_{N'}$
spontaneously broken to $U(1)_{1} \otimes ... \otimes U(1)_{M}$ with $M\leq N+N'$
is taken care of by replacing $(gg^T)^{-1}$  in \eq{2LU1match}
with the block-diagonal $(N+N')\times (N+N')$ matrix
\beq
(G G^T)^{-1} = \mbox{Diag}\left[(g g^T)^{-1},g_p^{-2} - \frac{C_2(G_p)}{48\pi^2} \right]
\label{2Lmatchgen}
\eeq
thus providing, together with the extended \eq{U1charges}
and \eq{2LU1match}, a generalization of \eq{2LmatchNOmix}.

\section{Numerical results}
\label{sec:results}

At one-loop, and in absence of the $U(1)$ mixing, the gauge
RGEs are not coupled and the unification constraints can be studied analytically.
When two-loop effects are included (or at one-loop more than one $U(1)$ factor is present) there is no closed solution and one must solve the system of coupled equations, matching all stages between the weak and unification scales, numerically.
On the other hand (when no $U(1)$ mixing is there)
one may take advantage of the analytic formula in \eq{alpha2loops}.
The latter turns out to provide, for the cases here studied, a very good approximation to the numerical solution. The discrepancies with the numerical integration do not
generally exceed the $10^{-3}$ level.

We perform a scan over the relevant breaking scales $M_{U}$, $M_{2}$ and $M_{1}$ and the value of the grand unified coupling $\alpha_{U}$ and
impose the matching with the SM gauge couplings at the $M_Z$ scale
requiring a precision at the per mil level.
This is achieved by minimizing the parameter
\beq
\delta=\sqrt{\sum_{i=1}^{3}\left(\frac{\alpha_i^{\rm th} -\alpha_i}{\alpha_i}\right)^2} \ ,
\label{deltaMZ}
\eeq
where $\alpha_i$ denote the experimental values at $M_Z$
and $\alpha_i^{\rm th}$ are the renormalized couplings obtained from unification.

The input values for the (consistently normalized) gauge SM couplings
at the scale $M_Z=91.19$ GeV are~\cite{PDG}
\bea
\alpha_1 &=& 0.016946 \pm 0.000006\,, \nn \\
\alpha_2 &=& 0.033812 \pm 0.000021\,, \\
\alpha_3 &=& 0.1176 \pm 0.0020\,,  \nn
\label{alphainormMZ}
\eea
corresponding to the electroweak scale parameters
\bea
\alpha^{-1}_{em} &=& 127.925 \pm 0.016\,, \nn \\
\sin^2\theta_W &=& 0.23119 \pm 0.00014\,.
\label{alphasinthetaMZ}
\eea
All these data refer to the modified minimally subtracted ($\overline{\mbox{MS}}$) quantities
at the $M_Z$ scale.

For $\alpha_{1,2}$ we shall consider only the central values while we resort to scanning over the whole $3\sigma$  domain for $\alpha_3$ when a stable solution is not found.

The results, i.e. the positions of the intermediate scales $M_{1}$, $M_{2}$ and $M_{U}$ shall be reported in terms of decadic logarithms of their values in units of GeV, i.e.
$n_1 = \log_{10}({M_1}/{\text{GeV}})$,
$n_2 = \log_{10}({M_2}/{\text{GeV}})$,
$n_U = \log_{10}({M_U}/{\text{GeV}})$.
In particular, $n_U$, $n_2$
are given as functions of $n_1$ for each breaking pattern
and for different approximations in the loop expansion. Each of the breaking patterns is further supplemented by the relevant range of the values of $\alpha_U$.

\subsection{$U(1)_R\times U(1)_X$ mixing}
\label{sec:U1RU1Xmixing}

The chains VIII to XII
require consideration of the mixing between the two $U(1)$ factors.
While $U(1)_{R}$ and $U(1)_{X}$ do emerge
with canonical diagonal kinetic
terms, being the remnants of the breaking of non-abelian groups,
the corresponding gauge couplings are at the onset different in size.
In general, no {\em scale independent}
orthogonal rotations of charges and gauge fields exist that diagonalize
the gauge interactions to all orders along the RGE trajectories.
According to the discussion in \sect{sec:2Lrge}, off-diagonal gauge
couplings arise at the one-loop level that must be accounted for
in order to perform the matching at the $M_1$ scale with
the standard hypercharge. The preserved direction in the $Q^{R,X}$ charge
space is given by
\beq
Q^Y = \sqrt{\frac{3}{5}} Q^R + \sqrt{\frac{2}{5}} Q^X\,,
\label{QYdef}
\eeq
where
\begin{equation}
Q^R = I_{3R}\;\;\; \text{ and } \;\;\;
Q^X = \sqrt{\frac{3}{2}}\ \left(\frac{B-L}{2}\right)\,.
\label{QRQXdefs}
\end{equation}
The matching of the gauge couplings is then obtained from \eq{2LU1match}
\beq
g_Y^{-2} = P\left(g g^T\right)^{-1} P^T\,,
\label{g_Ymatch}
\eeq
with
\beq
P = \left(\sqrt{\frac{3}{5}},\ \sqrt{\frac{2}{5}} \right)
\label{P_RX}
\eeq
and
\beq
g=
\left(\begin{array}{cc}
g_{RR}&g_{RX}\\[1.1ex]
g_{XR}&g_{XX}\\
\end{array}\right)\,.
\label{g_RX}
\eeq


When neglecting the off-diagonal terms, \eq{g_Ymatch} reproduces the matching condition
used in Refs.~\cite{Gipson:1984aj,Chang:1984qr,Deshpande:1992au,Deshpande:1992em}.
For all other cases, in which only one $U(1)$ factor is present,
the matching relations can be
read off directly from \eq{2LmatchGUT} and \eq{2LmatchNOmix}.

\subsection{Two-loop results (purely gauge)}
\label{sec:2loopgauge}

\begin{figure*}
 \centering
 \subfigure[\ Chain Ia]
   {\includegraphics[width=5.4cm]{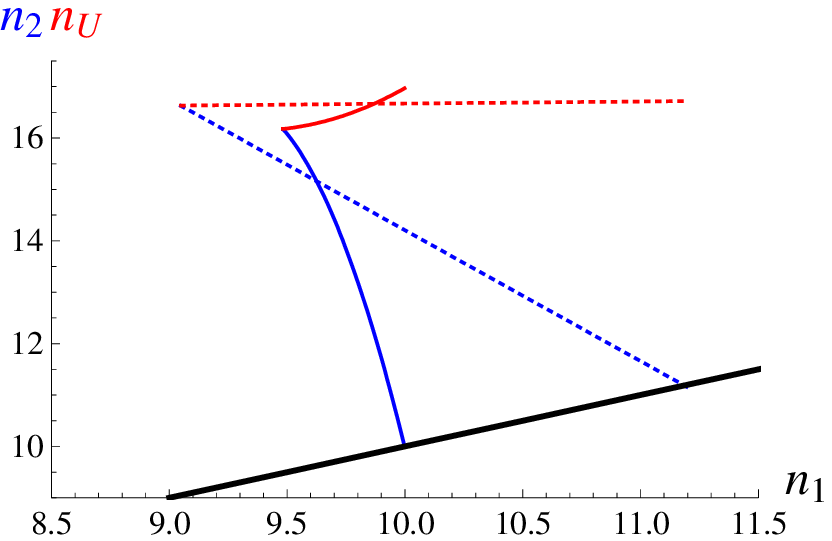}}
 \vspace{2mm}
 \subfigure[\ Chain IIa]
   {\includegraphics[width=5.4cm]{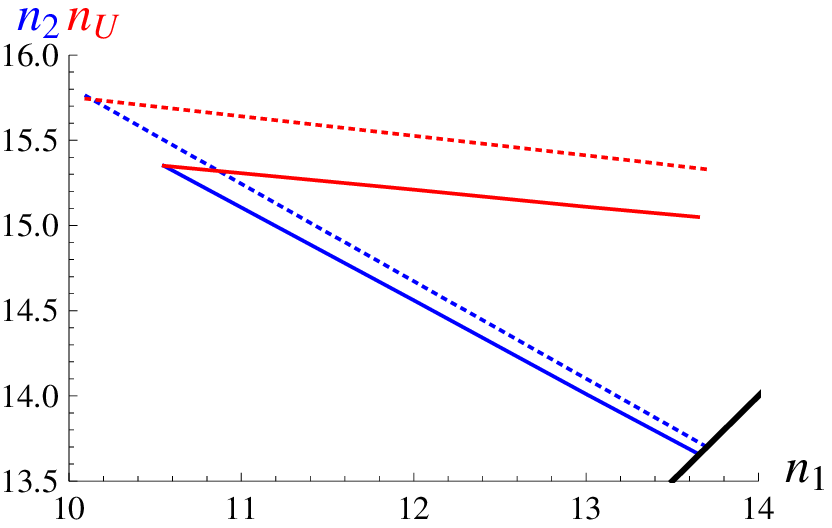}}
 \vspace{2mm}
 \subfigure[\ Chain IIIa]
   {\includegraphics[width=5.4cm]{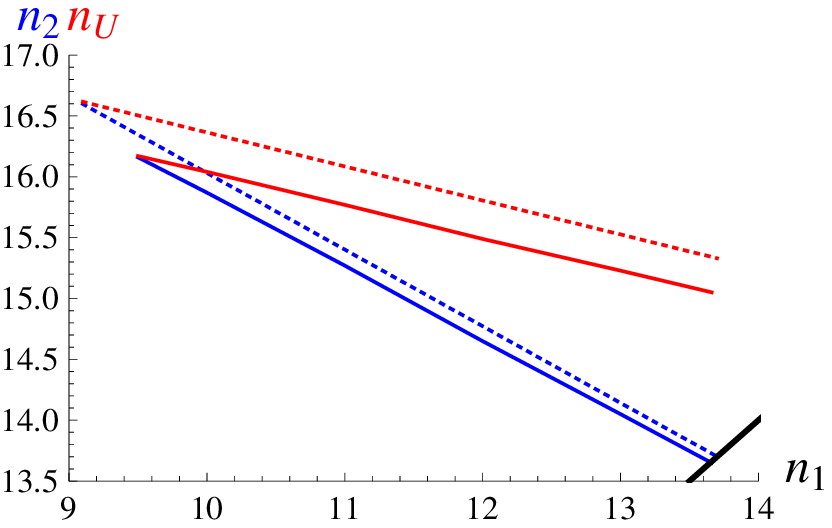}}
 \vspace{2mm}
\subfigure[\ Chain IVa]
   {\includegraphics[width=5.4cm]{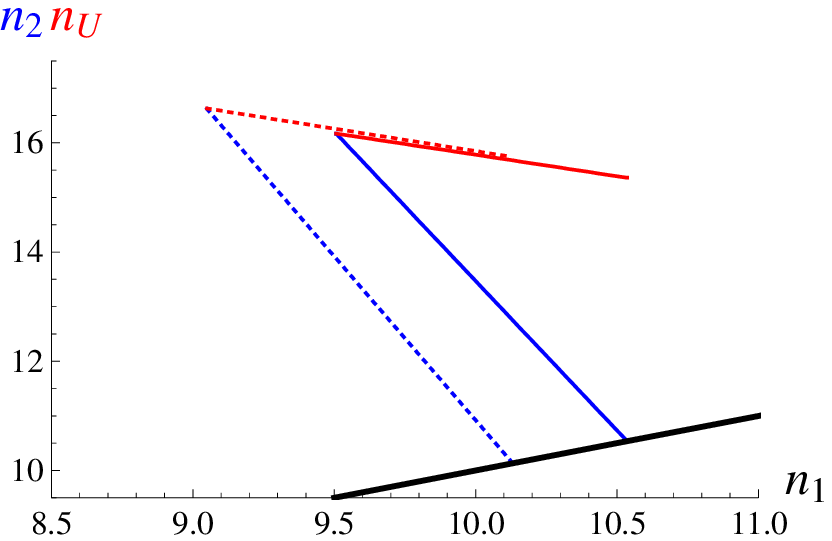}}
 \vspace{2mm}
 \subfigure[\ Chain Va]
   {\includegraphics[width=5.4cm]{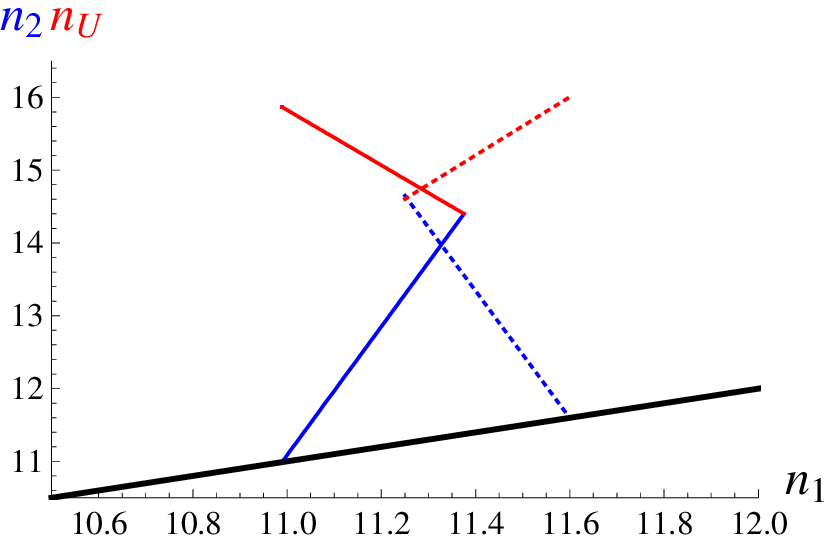}}
 \vspace{2mm}
 \subfigure[\ Chain VIa]
   {\includegraphics[width=5.4cm]{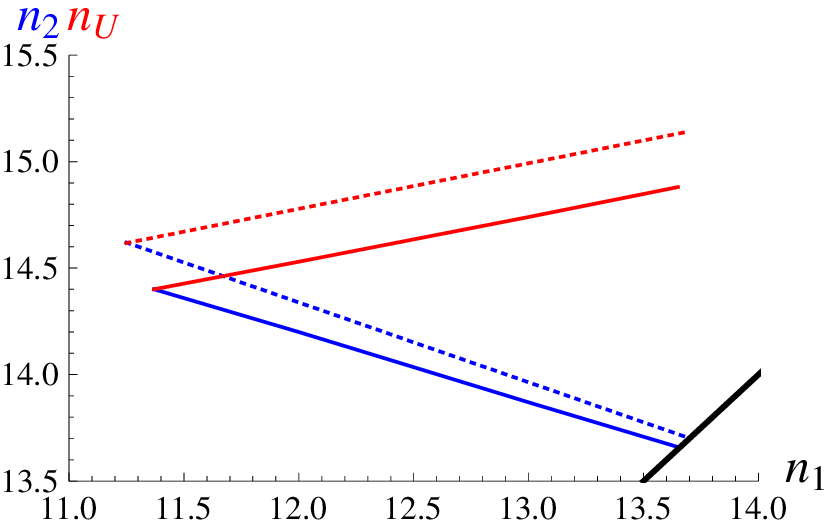}}
 \vspace{2mm}
 \subfigure[\ Chain VIIa]
   {\includegraphics[width=5.4cm]{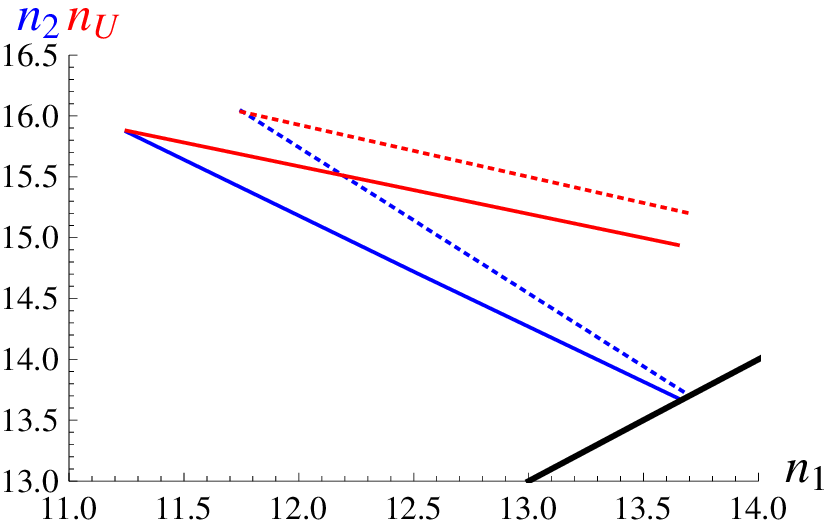}}
 \vspace{2mm}
 \subfigure[\ Chain VIIIa]
   {\includegraphics[width=5.4cm]{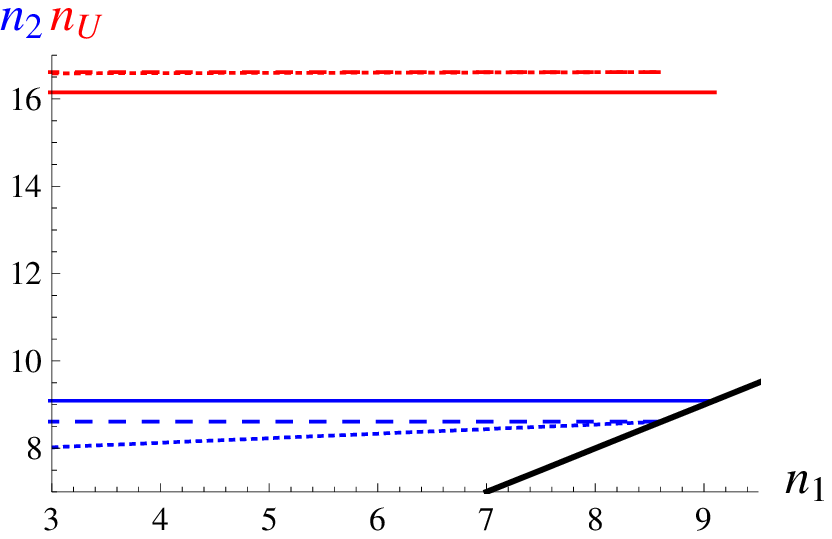}}
 \vspace{2mm}
 \subfigure[\ Chain IXa]
   {\includegraphics[width=5.4cm]{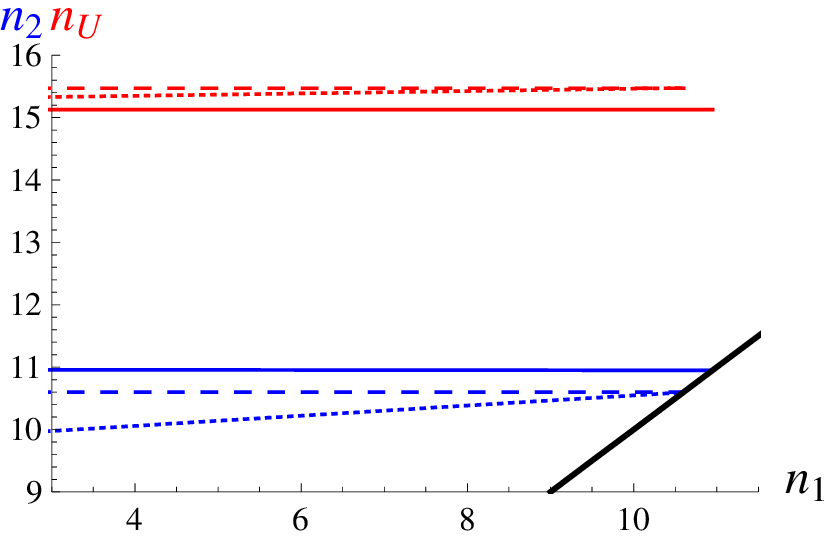}}
 \vspace{2mm}
 \subfigure[\ Chain XIa]
   {\includegraphics[width=5.4cm]{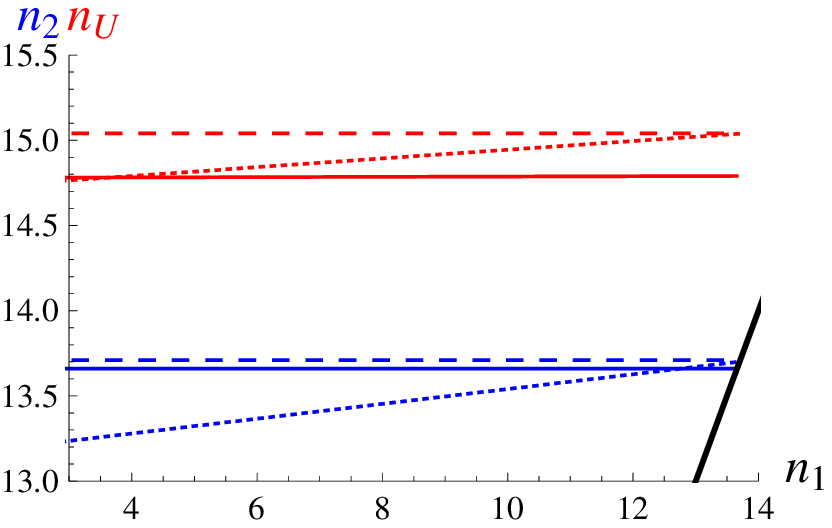}}
 \vspace{2mm}
 \subfigure[\ Chain XIIa]
   {\includegraphics[width=5.4cm]{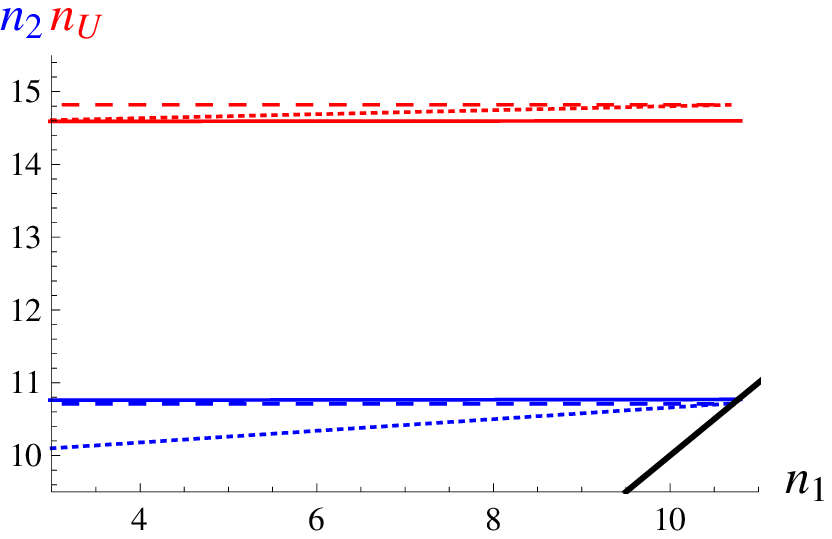}}
 \vspace{2mm}
\mycaption{The values of $n_U$ (red/upper branches)
and $n_2$ (blue/lower branches) are shown as functions
of $n_1$ for the pure gauge running in the $\overline{126}_H$ case.
The bold black line bounds the region $n_1 \leq n_2$.
From chains Ia to VIIa the short-dashed lines represent the result of one-loop running
while the solid ones correspond to the two-loop solutions. For chains
VIIIa to XIIa  the short-dashed lines represent the one-loop results without the $U(1)_{X}\otimes U(1)_{R}$ mixing,
the long-dashed lines account for the complete one-loop results,
while the solid lines represent the two-loop solutions.
The scalar content at each stage corresponds to that
considered in Ref.~\cite{Deshpande:1992em},
namely to that reported in \Table{tab:submultiplets} without the $\phi^{126}$ multiplets.
For chains I to VII the two-step $SO(10)$ breaking consistent with minimal fine tuning is recovered in the $n_2 \to n_U$ limit. No solution is found for chain Xa.}
\label{fig:1to12a}
\end{figure*}

\begin{figure*}
 \centering
 \subfigure[\ Chain Ib]
   {\includegraphics[width=5.4cm]{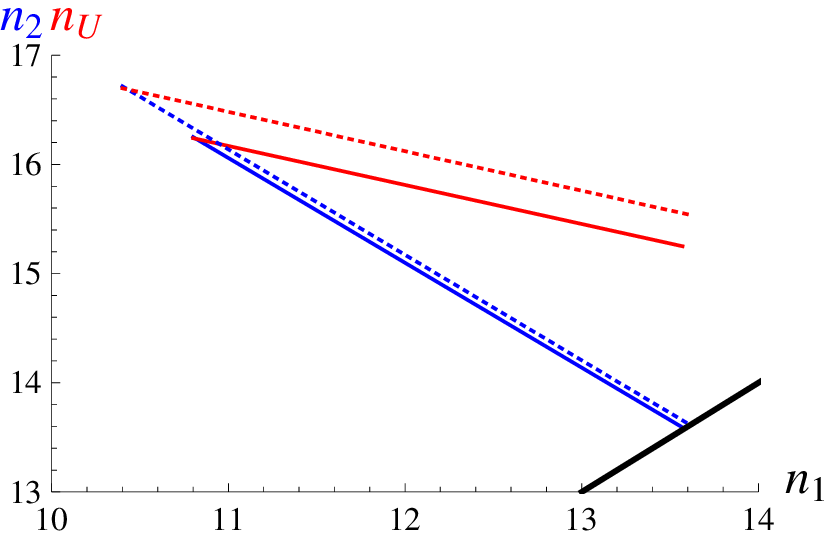}}
 \vspace{2mm}
 \subfigure[\ Chain IIb]
   {\includegraphics[width=5.4cm]{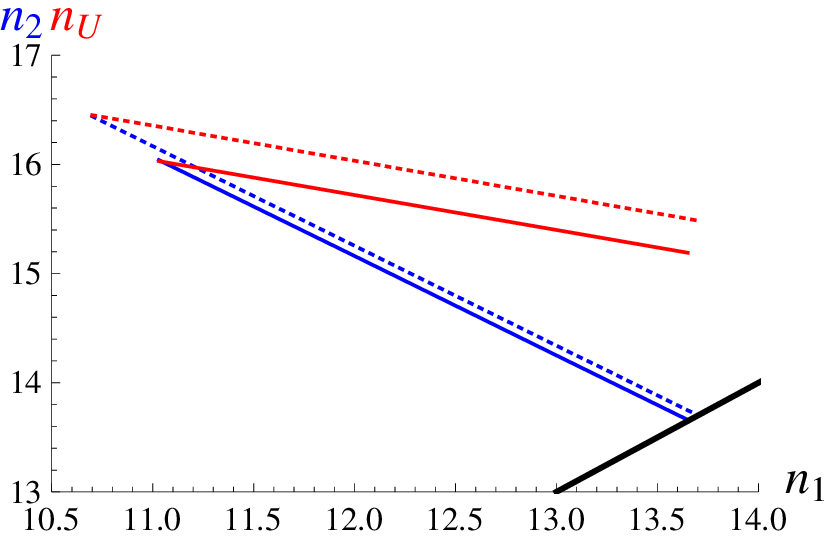}}
 \vspace{2mm}
 \subfigure[\ Chain IIIb]
   {\includegraphics[width=5.4cm]{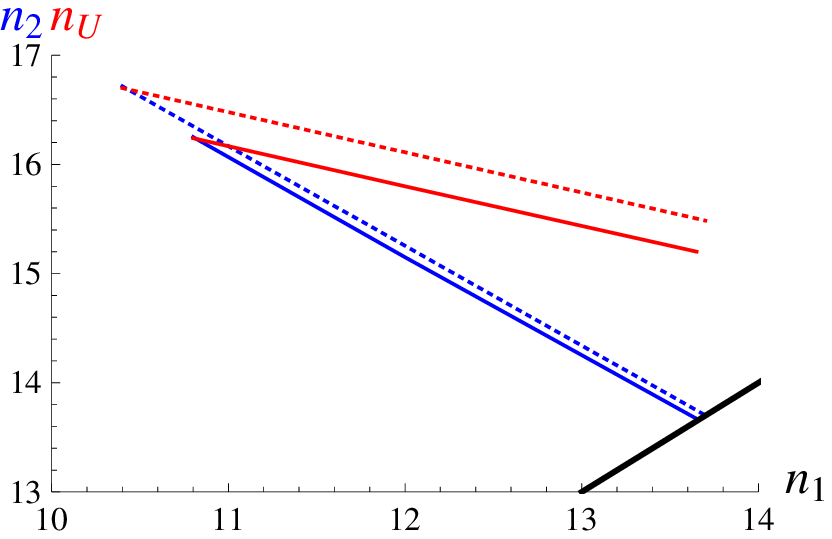}}
 \vspace{2mm}
\subfigure[\ Chain IVb]
   {\includegraphics[width=5.4cm]{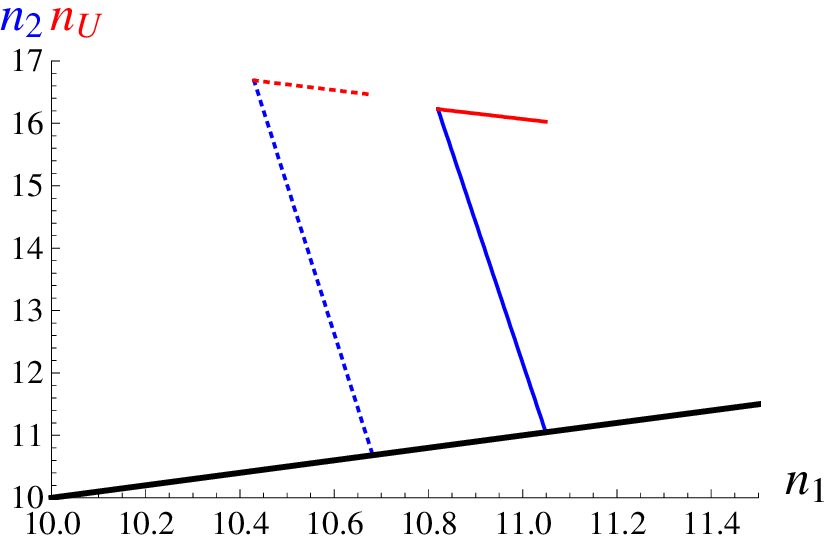}}
 \vspace{2mm}
 \subfigure[\ Chain Vb]
   {\includegraphics[width=5.4cm]{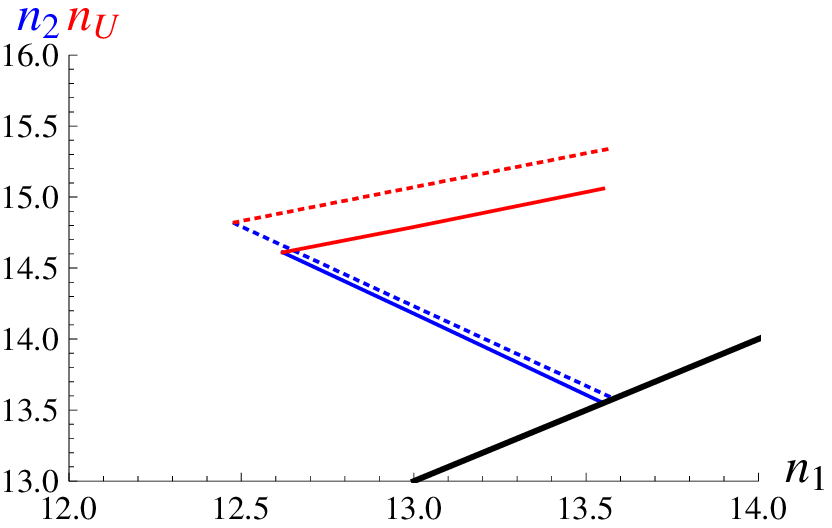}}
 \vspace{2mm}
 \subfigure[\ Chain VIb]
   {\includegraphics[width=5.4cm]{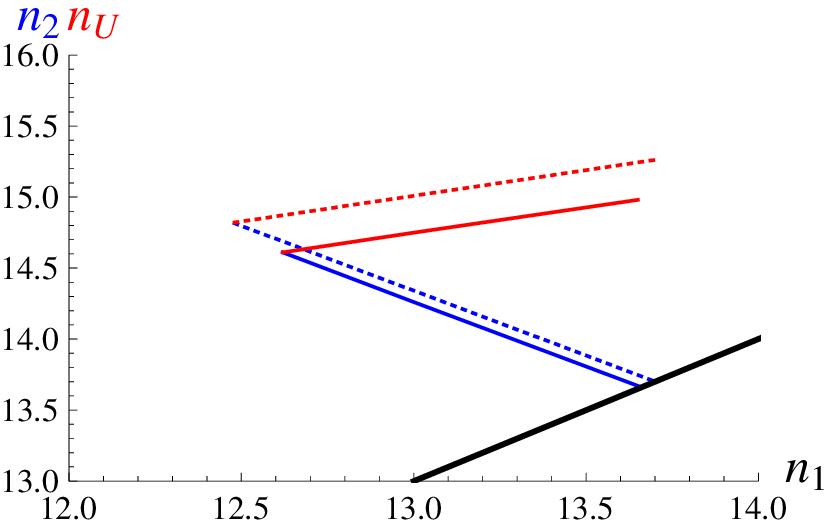}}
 \vspace{2mm}
 \subfigure[\ Chain VIIb]
   {\includegraphics[width=5.4cm]{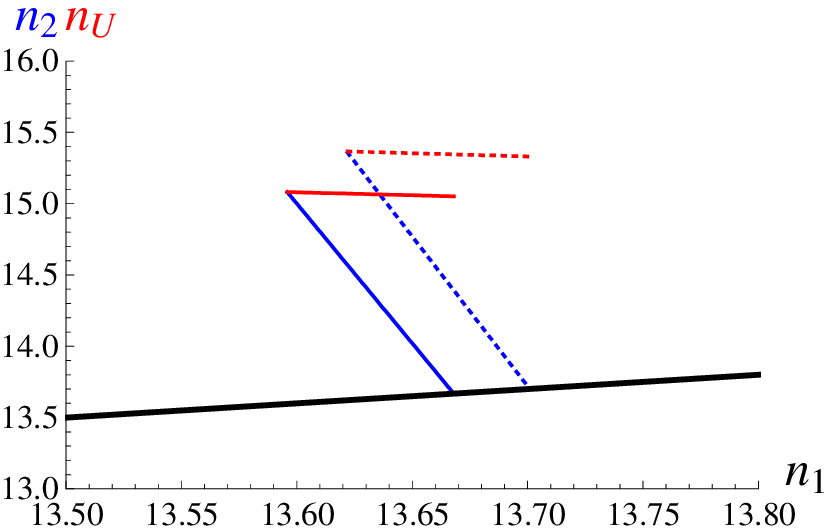}}
 \vspace{2mm}
 \subfigure[\ Chain VIIIb]
   {\includegraphics[width=5.4cm]{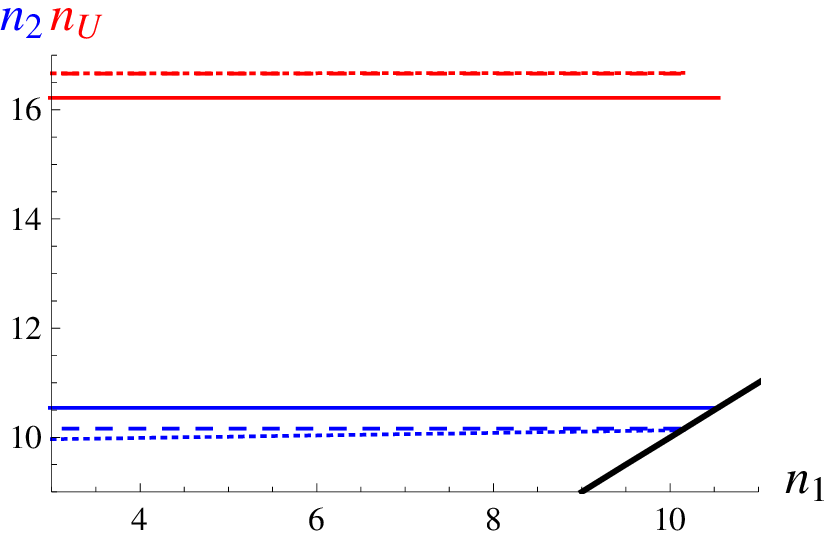}}
 \vspace{2mm}
 \subfigure[\ Chain IXb]
   {\includegraphics[width=5.4cm]{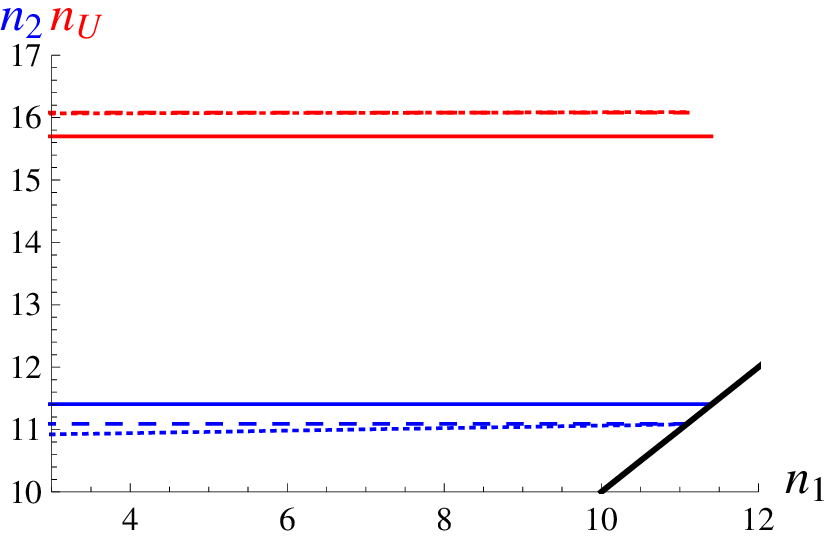}}
 \vspace{2mm}
\subfigure[\ Chain Xb]
   {\includegraphics[width=5.4cm]{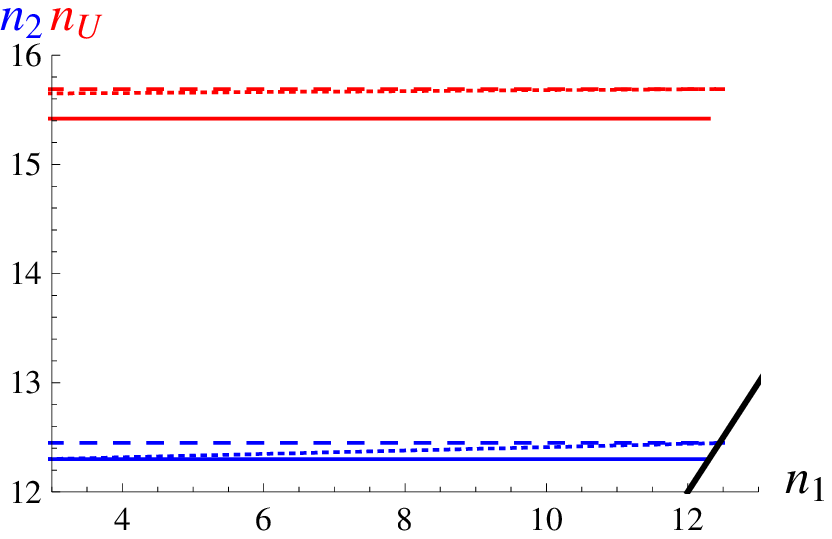}}
 \vspace{2mm}
 \subfigure[\ Chain XIb]
   {\includegraphics[width=5.4cm]{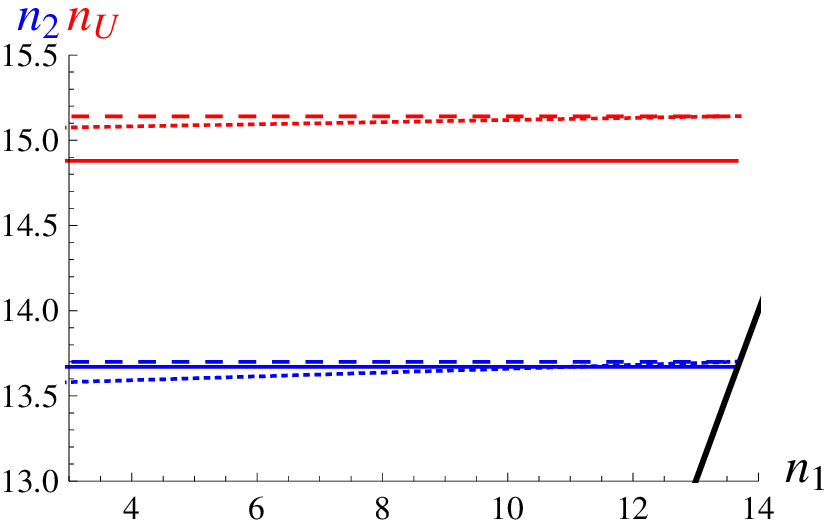}}
 \vspace{2mm}
 \subfigure[\ Chain XIIb]
   {\includegraphics[width=5.4cm]{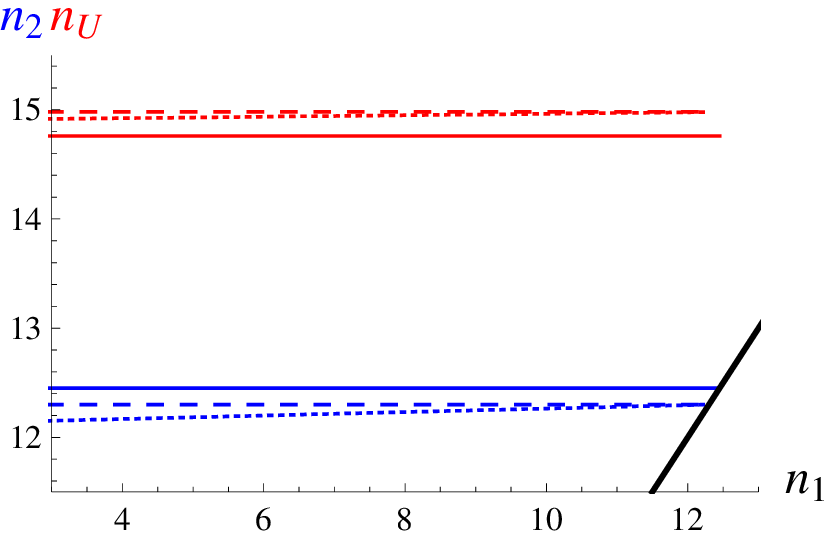}}
 \vspace{2mm}
\mycaption{Same as in \fig{fig:1to12a} for the $\overline{16}_H$ case.
}
\label{fig:1to12b}
\end{figure*}

The results of the numerical analysis are organized as follows:
\fig{fig:1to12a} and \fig{fig:1to12b} show the values of $n_U$ and $n_2$
as functions of $n_1$ for the pure gauge running (i.e. no Yukawa interactions),
in the $\overline{126}_H$ and $\overline{16}_H$ case respectively. The differences between the patterns for the $\overline{126}_H$ and
$\overline{16}_H$ setups depend on the substantially different scalar content.
The shape and size of the various contributions (one-loop, with and without $U(1)$ mixing,
and two-loops) are compared in each figure.
The dissection of the RGE results shown in the figures allows us
to compare our results with those
of Refs.~\cite{Gipson:1984aj,Chang:1984qr,Deshpande:1992au,Deshpande:1992em}.

\Table{tab:alphaU} shows the two-loop values of $\alpha_U^{-1}$ in the
allowed region for $n_1$.
The contributions of the additional $\phi^{126}$ multiplets,
and the Yukawa terms are discussed separately in \sect{sec:extrahiggs}
and \sect{sec:yukawaterms}, respectively.
With the exception of a few singular cases detailed therein, these effects turn out to be generally subdominant.

As already mentioned in the introduction, two-loop precision
in a GUT scenario makes sense once (one-loop) thresholds effects are coherently
taken into account, as their effect may become comparable if not
larger than the two loop itself (the argument becomes stronger as the number
of intermediate scales increases). On the other hand, there is no control
on the spectrum unless a specific model is studied in details.
The purpose of this work is to set the stage for such a study
by reassessing and updating the general constraints and patterns
that $SO(10)$ grand unification enforces on the spread of intermediate
scales.

The one and two-loop $\beta$-coefficients used in the present study are
reported in Appendix \ref{app:2Lbeta}. \Table{tab:betaChang} in the appendix shows
the reduced $\widetilde b_{ij}$ coefficients for those cases where we are at variance
with Ref.~\cite{Chang:1984qr}.

One of the largest effects in the comparison with
Refs.~\cite{Gipson:1984aj,Chang:1984qr,Deshpande:1992au,Deshpande:1992em}
emerges at one-loop and it is due to the implementation of the $U(1)$ gauge mixing
when $U(1)_R \otimes U(1)_X$ appears as an intermediate stage of the $SO(10)$
breaking\footnote{The lack of abelian gauge mixing in Ref.~\cite{Deshpande:1992em}
was first observed in Ref.~\cite{Lavoura:1993ut}.}.
This affects chains VIII to XII, and it exhibits itself in the exact (one-loop)
flatness of $n_2$, $n_U$ and $\alpha_U$ as functions of $n_1$.

The rationale for such a behaviour is quite simple.
When considering
the gauge coupling renormalization in the $2_L 1_R 1_X 3_c$ stage,
no effect at one-loop appears in the non-abelian $\beta$-functions
due to the abelian gauge fields. On the other hand,
the Higgs fields surviving at the $2_L 1_R 1_X 3_c$ stage, responsible
for the breaking to $1_Y 2_L 3_c$, are (by construction) SM singlets.
Since the SM one-loop $\beta$-functions are not affected by their presence,
the solution found for $n_2$, $n_U$ and $\alpha_U$ in the $n_1 = n_2$ case
holds for $n_1 < n_2$ as well.
Only by performing correctly the mixed $1_R 1_X$
gauge running and the consistent matching with $1_Y$
one recovers the expected $n_1$ flatness of the GUT solution.

In this respect, it is interesting to notice that the absence of $U(1)$ mixing in
Refs.~\cite{Gipson:1984aj,Chang:1984qr,Deshpande:1992au,Deshpande:1992em}
makes the argument for the actual possibility of a light (observable) $U(1)_R$ gauge boson
an ``approximate" statement (based on the approximate flatness of the solution).

One expects this feature to break at two-loops. The $SU(2)_L$ and $SU(3)_c$ $\beta$-functions
are affected at two-loops directly by the abelian gauge bosons via \eq{bpU12loops}
(the Higgs multiplets
that are responsible for the $U(1)_R\otimes U(1)_X$ breaking do not enter through the Yukawa interactions).
The net effect on the non-abelian gauge running is related to the difference between the
contribution of the $U(1)_R$ and $U(1)_X$ gauge bosons  and that of the standard hypercharge.
We checked that such a difference is always a small fraction (below 10\%) of the
typical two-loop contributions to the $SU(2)_L$ and $SU(3)_c$  $\beta$-functions.
As a consequence, the $n_1$ flatness of the GUT solution is at a very high accuracy
($10^{-3}$) preserved at two-loops as well, as the inspection of the relevant chains
in \figs{fig:1to12a}{fig:1to12b} shows.

Still at one-loop we find a sharp disagreement in the $n_1$ range of chain XIIa,
with respect to the result of Ref.~\cite{Deshpande:1992em}.
The authors find $n_1 < 5.3$, while
strictly following their procedure and assumptions we find $n_1 < 10.2$
(the updated one- and two-loop results are given in \fig{fig:1to12a}k).
As we shall see, this difference brings chain XIIa back among the potentially realistic ones.

\begin{table}
\begin{tabular}{lcclc}
\hline
{\rm Chain}  &  {\rm $\alpha^{-1}_U$}   && {\rm Chain}  &  {\rm $\alpha^{-1}_U$}  \\
\hline
Ia  & $[45.5,46.4]$ && Ib  & $[45.7,44.8]$
\\
IIa  & $[43.7,40.8]$ && IIb  & $[45.3,44.5]$
\\
IIIa  & $[45.5,{40.8}]$ && IIIb  & $[45.7,{44.5}]$
\\
IVa  & $[45.5,43.4]$ && IVb  & $[45.7,45.1]$
\\
Va  & $[45.4,44.1]$ && Vb  & $[44.3,44.8]$
\\
VIa  & $[44.1,41.0]$ && VIb  & $[44.3,44.2]$
\\
VIIa  & $[45.4,41.1]$ && VIIb  & $[44.8,44.4]$
\\
VIIIa  & $45.4$ && VIIIb  & $45.6$
\\
IXa  & $42.8$ && IXb  & $44.3$
\\
Xa  & $ $ && Xb  & $44.8$
\\
XIa  & $38.7$ && XIb  & $41.5$
\\
XIIa  & $44.1$ && XIIb  & $44.3$
\\
\hline
\end{tabular}

\mycaption{Two-loop values of $\alpha^{-1}_U$ in the allowed region for $n_1$.
From chains I to VII, $\alpha^{-1}_U$ is $n_1$ dependent and its range is given in square brackets for the minimum (left) and the maximum (right) value of $n_1$ respectively.
For chains VIII to XII, $\alpha^{-1}_U$ depends very weekly on $n_1$ (see the discussion on $U(1)$ mixing in the text). No solution is found for chain Xa.}
\label{tab:alphaU}
\end{table}

As far as two-loop effects are at stakes, their relevance is generally related to the
length of the running involving the largest non-abelian groups.
On the other hand, there are chains where $n_2$ and $n_U$ have a strong dependence
on $n_1$ (we will refer to them as to ``unstable" chains) and where two-loop
corrections affect substantially the one-loop results.
Evident examples of such unstable chains are Ia, IVa, Va, IVb, and VIIb.
In particular, in chain Va the two-loop effects flip
the slopes of $n_{2}$ and $n_{U}$, that implies a sharp change in the allowed region for $n_1$.
It is clear that when dealing with these breaking chains any
statement about their viability should account for the details of the thresholds
in the given model.

In chains VIII to XII (where the second intermediate stage is $2_L 1_R 1_X 3_c$,
two-loop effects are mild
and exhibit the common behaviour
of lowering the GUT scale ($n_U$) while raising (with the exception of Xb and XIa,b)
the largest intermediate scale ($n_2$).
The mildness of two-loop corrections (no more that one would a-priori expect) is strictly
related to the ($n_1$) flatness of the GUT solution discussed before.

Worth mentioning are the limits $n_2\sim n_U$ and $n_1\sim n_2$.
While the former is equivalent to neglecting the first stage $G2$
and to reducing effectively the three breaking steps to just two
(namely $SO(10)\rightarrow G1\rightarrow SM$) with a minimal fine tuning in the scalar sector, care must be taken of the latter.
One may naively expect that the chains with the same G2
should exhibit for $n_1\sim n_2$ the same numerical behavior ($SO(10)\rightarrow G2\rightarrow SM$), thus
clustering the chains (I,V,X), (II,III,VI,VII,XI) and (IV,IX).
On the other hand, one must recall that the existence of G1 and its breaking
remain encoded in the G2 stage through the Higgs scalars that are responsible
for the G2$\to$G1 breaking. This is why the chains with the same G2 are
not in general equivalent in the $n_1\sim n_2$ limit.
The numerical features of the degenerate patterns (with $n_2\sim n_U$) can be
crosschecked among the different chains by direct
inspection of \figs{fig:1to12a}{fig:1to12b} and \Table{tab:alphaU}.

In any discussion of viability of the various scenarios the main attention is paid to
the constraints emerging from the proton decay.
In non supersymmetric GUTs, this process is mediated by baryon number violating
gauge interactions, inducing at low energies
a set of effective dimension 6 operators that conserve $B-L$.
In the $SO(10)$ scenarios we consider here,
such gauge bosons are integrated out at the unification scale, and therefore
proton decay constrains $n_U$ from below. The present experimental limit
$\tau_p(p\rightarrow e^+\pi^0) > 1.6\times 10^{33}$ years~\cite{PDG} implies
\beq
\left(\frac{\alpha^{-1}_U}{45}\right) 10^{2(n_U-15)} > 5.2\,,
\label{pdecaybound}
\eeq
that, for $\alpha^{-1}_U = 45$ yields $n_U > 15.4$.
Taking the results in \figs{fig:1to12a}{fig:1to12b} and \Table{tab:alphaU} at face value the chains VIab, XIab, XIIab, Vb and VIIb should be excluded from realistic considerations.

On the other hand, one must recall that once a specific model is scrutinized
in detail there can be large threshold corrections in the
matching~\cite{Dixit:1989ff,Mohapatra:1992dx,Lavoura:1993su},
that can easily move the unification scale by a few orders of magnitude (in both directions).
In particular, as a consequence of the spontaneous breaking of accidental would-be
global symmetries of the scalar
potential, pseudo-Goldstone modes (with
masses further suppressed with respect to the expected threshold range) may appear in the scalar spectrum, leading to
potentially large RGE effects~\cite{Aulakh:1982sw}.
Therefore, we shall follow a conservative approach in interpreting the limits on the
intermediate scales coming from a simple threshold clustering. These limits, albeit
useful for a preliminary survey, may not be sharply used to exclude marginal
but otherwise well motivated scenarios.

Below the scale of the $B-L$ breaking, processes that violate separately the barion or the lepton
numbers emerge.
In particular,
$\Delta B = 2$ effective interactions give rise to the phenomenon of neutron oscillations
(for a recent review see Ref.~\cite{Mohapatra:2009wp}).
Present bounds on nuclear instability give $\tau_{Nucl}> 10^{32}$ years, which
translates into a bound on the neutron oscillation time $\tau_{n-\bar n} > 10^8$ sec.
Analogous limits come from direct reactor oscillations experiments.
This sets a lower bound on the scale of $\Delta B = 2$ non supersymmetric (dimension 9) operators
that varies from 10 to 300 TeV depending on model couplings.
Thus, neutron-antineutron
oscillations probe scales far below the unification scale. In a supersymmetric
context the presence of $\Delta B = 2$
dimension 7 operators softens the dependence on the $B-L$ scale and for the present bounds
the typical limit goes up to about $10^7$ GeV.

Far more reaching in scale sensitivity are the $\Delta L = 2$
neutrino masses emerging from the see-saw mechanism.
At the $B-L$ breaking scale the $\Delta^{126}_R$ ($\delta^{16}_R$) scalars
acquire $\Delta L = 2$ ($\Delta L = 1$) vacuum expectation values (VEVs)
that give a Majorana mass to the right-handed neutrinos. Once the latter are integrated out,
dimension 5 operators of the form $\bar \nu^c_L \nu_L H H^T$
generate light Majorana neutrino states in the low
energy theory.

In the type-I seesaw,  the neutrino mass matrix $m_{\nu}$ is proportional
to $Y_N M_R^{-1} Y_N^Tv^{2}$ where the largest entry
in the Yukawa couplings is typically of the order of the top quark one and $M_{R}\sim M_{1}$.
Given a neutrino mass above the limit obtained from atmospheric neutrino oscillations and below the eV, one infers a (loose) range $10^{12}\ \mbox{GeV} < M_1 < 10^{14}\ \mbox{GeV}$.
It is interesting to note that the lower bound pairs with the cosmological limit
on the D-parity breaking scale (see \sect{sec:chains}).

In the scalar-triplet induced (type-II) seesaw the evidence of the neutrino mass
entails a lower bound on the VEV of the heavy $SU(2)_{L}$ triplet in $\overline{126}_{H}$
(or in $\overline{16}_{H}\overline{16}_{H}$). This translates into an upper bound on the
mass of the triplet that depends on the structure of the relevant Yukawa coupling.
If both type-I as well as type-II contribute to the light neutrino mass,
the lower bound on the $M_{1}$ scale may then be weakened by the interplay between
the two contributions. Once again this can be quantitatively assessed only when the
vacuum of the model is fully investigated.

Finally, it is worth noting that if the $B-L$ breakdown is driven by $\overline{126}_{H}$,
the elementary triplets couple to the Majorana currents at the renormalizable level and $m_{\nu}$
is directly sensitive to the position of the $G1\to SM$ threshold $M_{1}$.
On the other hand, the $n_{1}$-dependence of $m_{\nu}$ is loosened in the b-type of chains
due to the non-renormalizable nature of the relevant effective operator
$16_{F}16_{F}\overline{16}_{H}\overline{16}_{H}/\Lambda$, where the effective
scale $\Lambda>M_{U}$ accounts for an extra suppression.

With these considerations at hand, the constraints from proton decay and the see-saw
neutrino scale favor the chains II, III and VII, which all share $2_L 2_R 4_C P$
in the first $SO(10)$ breaking stage~\cite{Bajc:2005zf}.
On the other hand, our results rescue from oblivion
other potentially interesting scenarios
that, as we shall expand upon shortly, are worth of in depth consideration.
In all cases, the bounds on the $B-L$ scale enforced by the see-saw neutrino mass
excludes the possibility of observable $U(1)_R$ gauge bosons.

\subsection{The $\phi^{126}$ Higgs multiplets}
\label{sec:extrahiggs}

As mentioned in \sect{sec:ESH},
in order to ensure a rich enough Yukawa sector in realistic models
there may be the need to keep more than one $SU(2)_{L}$ Higgs doublet at
intermediate scales, albeit at the price of an extra fine-tuning.
A typical example is the
case of a relatively low Pati-Salam breaking scale where
one needs at least a pair of $SU(2)_{L}\otimes SU(2)_{R}$ bidoublets with different $SU(4)_{C}$ quantum numbers to
transfer the information about the PS breakdown into the matter sector. Such
additional Higgs multiplets are those labelled by $\phi^{126}$ in \Table{tab:submultiplets}.

\Table{tab:phi126chains} shows the effects of including $\phi^{126}$ at the
$SU(4)_{C}$ stages of the relevant breaking chains. The two-loop
results at the extreme values of the intermediate scales, with and without
the $\phi^{126}$ multiplet, are compared.
In the latter case the complete functional dependence among
the scales is given in \fig{fig:1to12a}. Degenerate patterns
with only one effective intermediate stage are easily crosschecked among the different chains in \Table{tab:phi126chains}.

In most of the cases, the numerical results do not exhibit a sizeable dependence on the additional $(2,2,15)_{\overline{126}}$ (or $(2,+\frac{1}{2},15)_{\overline{126}}$)
scalar multiplets. The reason can be read off
\Table{tab:phi126betas} in Appendix \ref{app:2Lbeta} and it rests on an
accidental approximate coincidence of the $\phi^{126}$ contributions to the $SU(4)_{C}$ and $SU(2)_{L,R}$ one-loop beta coefficients
(the same argument applies to the  $2_{L}1_{R}4_{C}$ case).

Considering for instance the $2_{L}2_{R}4_{C}$ stage, one obtains
$\Delta a_{4}=\frac{1}{3} \times 4\times T_{2}(15)=\frac{16}{3}$,
and $\Delta a_{2}=\frac{1}{3}\times 30 \times T_{2}(2)=5$, that only slightly affects
the value of $\alpha_U$ (when the PS scale is low enough), but has generally
a negligible effect on the intermediate scales.

\begin{table}
\begin{tabular}{lcccc}
\hline
{\rm Chain}     & {\rm $n_1$}  & {\rm $n_2$}  & {\rm $n_U$}  & {\rm $\alpha^{-1}_U$}
\\
\hline
Ia              & [9.50, 10.0] & [16.2, 10.0] & [16.2, 17.0] & [45.5, 46.4]
\\
                & [8.00, 9.50] & [10.4, 16.2] & [18.0, 16.2] & [30.6, {45.5}]
\\
IIa             & [10.5, 13.7] & [15.4, 13.7] & [15.4, 15.1] & [43.7, 40.8]
\\
                & [10.5, 13.7] & [15.4, 13.7] & [15.4, 15.1] & [43.7, 37.6]
\\
IIIa            & [9.50, 13.7] & [16.2, 13.7] & [16.2, 15.1] & [45.5, {40.8}]
\\
                & [9.50, 13.7] & [16.2, 13.7] & [16.2, 15.1] & [45.5, {37.6}]
\\
Va              & [11.0, 11.4] & [11.0, 14.4] & [15.9, 14.4] & [45.4, 44.1]
\\
                & [10.1, 11.2] & [10.1, 14.5] & [16.5, 14.5] & [32.5, 40.8]
\\
VIa             & [11.4, 13.7] & [14.4, 13.7] & [14.4, 14.9] & [44.1, 41.0]
\\
                & [11.2, 13.7] & [14.5, 13.7] & [14.5, 14.9] & [40.8, 38.1]
\\
VIIa            & [11.3, 13.7] & [15.9, 13.7] & [15.9, 14.9] & [45.4, 41.1]
\\
                & [10.5, 13.7] & [16.5, 13.7] & [16.5, 15.0] & [33.3, 38.1]
\\
XIa             & [3.00, 13.7] & [13.7, 13.7] & [14.8, 14.8] & [38.7, 38.7]
\\
                & [3.00, 13.7] & [13.7, 13.7] & [14.8, 14.8] & [36.0, 36.0]
\\
XIIa            & [3.00, 10.8] & [10.8, 10.8] & [14.6, 14.6] & [44.1, 44.1]
\\
                & [3.00, 10.5] & [10.5, 10.5] & [14.7, 14.7] & [39.8, 39.8]
\\

\hline
\end{tabular}
\mycaption{Impact of the additional multiplet $\phi^{126}$
(second line of each chain) on those chains that contain the gauge
groups $2_L2_R4_C$ or $2_L1_R4_C$ as intermediate stages,
and whose breaking to the SM is obtained via a $\overline{126}_H$ representation.
The values of $n_2$, $n_U$ and $\alpha^{-1}_U$ are showed for
the minimum and maximum values allowed for $n_1$ by the two-loop analysis.
Generally the effects on the intermediate scales
are below the percent level,
with the exception of chains Ia and Va that are most sensitive to
variations of the $\beta$-functions.}
\label{tab:phi126chains}
\end{table}

An exception to this argument is observed in chains
Ia and Va that, due to their $n_{2,U}(n_1)$ slopes, are most sensitive to variations of the $\beta$-coefficients. In particular, the inclusion of $\phi^{126}$ in the Ia chain flips at two-loops
the slopes of $n_2$ and $n_U$ so that the limit $n_2 = n_U$ (i.e. no G2 stage) is obtained for the maximal value of $n_1$ (while the same happens for the minimum $n_1$ if there is no $\phi^{126}$).

\begin{figure}
 \centering
 \subfigure[\ Chain Ia]
   {\includegraphics[width=5.4cm]{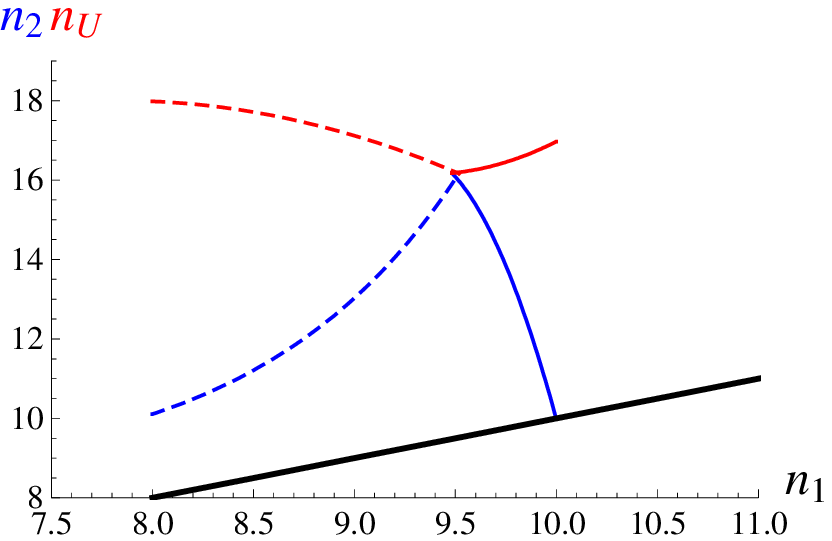}}
 \vspace{2mm}
 \subfigure[\ Chain Va]
   {\includegraphics[width=5.4cm]{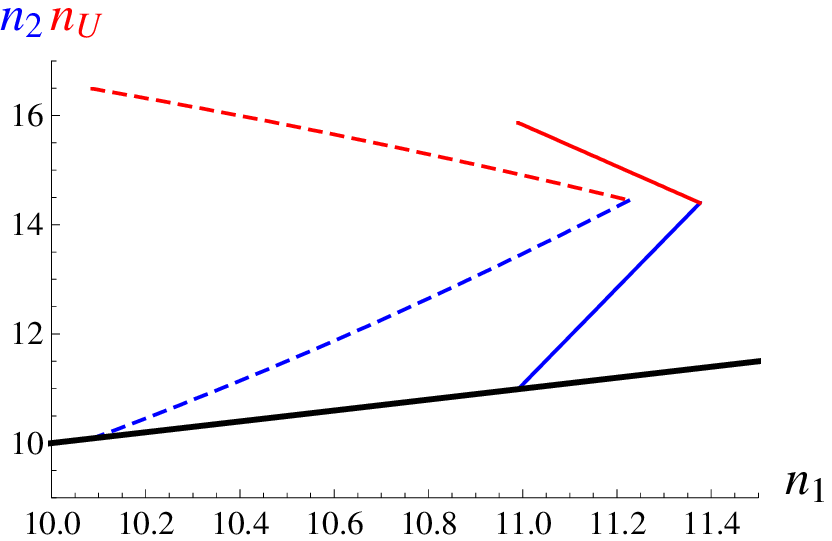}}
 \vspace{2mm}
 \subfigure[\ Chain VIIa]
   {\includegraphics[width=5.4cm]{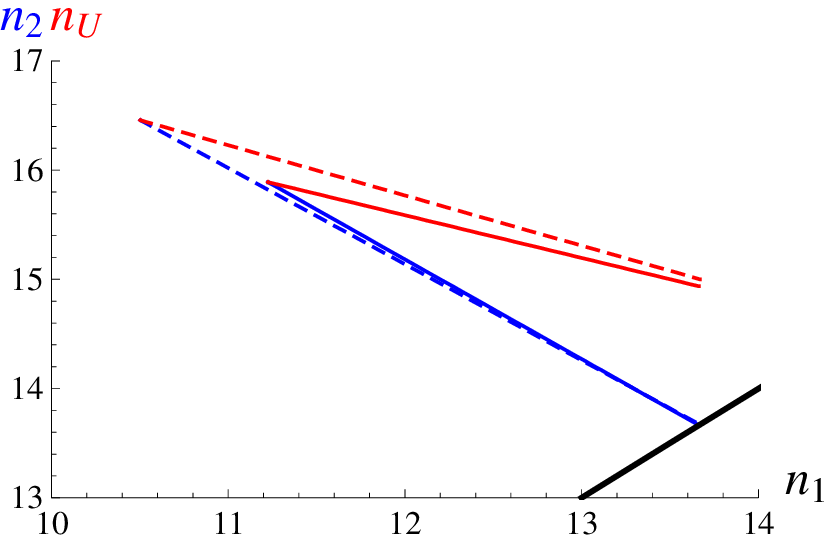}}
\mycaption{Example of chains with sizeable $\phi^{126}$ effects (long-dashed curves)
on the position of the intermediate scales.
The solid curves represent the two-loop results in \fig{fig:1to12a}.
The most dramatic effects appear in the chain Ia, while moderate scale shifts
affect chain Va (both ``unstable" under small variations of the $\beta$-functions).
Chain VIIa, due to the presence of two PS stages, is the only "stable''
chain with visible $\phi^{126}$ effects.
}
\label{fig:phi126}
\end{figure}

\fig{fig:phi126} shows three template cases where the $\phi^{126}$ effects are visible.
The highly unstable Chain Ia shows, as noticed earlier, the largest effects.
In chain Va the
effects of $\phi^{126}$ are moderate. Chain VII is the only "stable" chain that
exhibits visible effects on the intermediate scales.
This is due to the presence of two full-fledged PS stages.

\subsection{Yukawa terms}
\label{sec:yukawaterms}


The effects of the Yukawa couplings can be at leading order approximated by constant negative shifts of the one-loop $a_{i}$ coefficients $a_{i}\to a_{i}'=a_{i}+\Delta a_{i} $ with
\begin{equation}
\Delta a_{i}=-\frac{1}{(4\pi)^{2}}y_{ik}\text{Tr}\,Y_{k}\,Y_{k}^{\dagger}\,.
\end{equation}
The impact of $\Delta a_{i}$ on the position of the unification scale and the value of the unified coupling can be simply estimated by considering the running induced by the Yukawa couplings from a scale $t$ up to the unification point ($t=0$). The one-loop result for the change of the intersection of the curves corresponding to
$\alpha_{i}^{-1}(t)$ and $\alpha_{j}^{-1}(t)$ reads (at the leading order in
$\Delta a_{i}$):
\begin{equation}
\Delta t_{U}= 2\pi \frac{\Delta a_{i}-\Delta a_{j}}{(a_{i}-a_{j})^{2}}\left[\alpha_{j}^{-1}(t_{})-\alpha_{i}^{-1}(t_{})\right]+\ldots
\end{equation}
and
\begin{eqnarray}
\Delta \alpha_{U}^{-1}& =&  \frac{1}{2}\left[\frac{\Delta a_{i}+\Delta a_{j}}{a_{i}-a_{j}}-\frac{(a_{i}+a_{j})(\Delta a_{i}-\Delta a_{j})}{(a_{i}-a_{j})^{2}}\right]\nonumber\\[1ex]
&& \times\left[\alpha_{j}^{-1}(t_{})-\alpha_{i}^{-1}(t_{})\right]+\ldots
\label{eq:deltaalphaUinv}
\end{eqnarray}
for any $i\neq j$.
For simplicity we have neglected the changes in the $a_{i}$ coefficients due to crossing intermediate thresholds.
It is clear that for a common change $\Delta a_{i}=\Delta a_j$ the unification scale is not affected,
while a net effect remains on $\alpha_{U}^{-1}$.
In all cases, the leading contribution is always proportional to
$\alpha_{j}^{-1}(t_{})-\alpha_{i}^{-1}(t_{})$ (this holds exactly for
$\Delta t_U$).

In order to assess quantitatively such effects we shall consider first the SM stage that accounts for a large part of the running in all realistic chains.
The case of a low $n_{1}$ scale
leads, as we explain in the following, to comparably smaller effects.
The impact of the Yukawa interactions on the gauge RGEs is readily estimated assuming only the up-type Yukawa contribution to be sizeable and constant,
namely $\text{Tr}\,Y_{U}\,Y_{U}^{\dagger}\sim 1$.
This yields $\Delta a_{i}\sim -6\times 10^{-3}y_{iU}$, where
the values of the $y_{iU}$ coefficients are given in Table~\ref{tab:Yukawas}.
For $i=1$ and $j=2$ one obtains $\Delta a_{1}\sim -1.1\times 10^{-2}$ and $\Delta a_{2}\sim -0.9\times 10^{-2}$ respectively.
Since $a_{1}^{SM}=\frac{41}{10}$ and $a_{2}^{SM}=-\frac{19}{6}$, the first term in (\ref{eq:deltaalphaUinv}) dominates and one finds $\Delta \alpha_{U}^{-1}\sim 0.04$.
For a typical value of $\alpha_{U}^{-1}\sim 40$ this translates into $\Delta \alpha_{U}^{-1}/\alpha_{U}^{-1}\sim 0.1\%$.
The impact on $t_U$ is indeed tiny, namely $\Delta n_{U}\sim -1\times 10^{-2}$.
In both cases the estimated effect agrees to high accuracy
with the actual numerical behavior we observe.

The effects of the Yukawa interactions emerging at intermediate scales (or of a non-negligible $\text{Tr }Y_{D}\,Y_{D}^{\dagger}$ in a two Higgs doublet settings with large $\tan\beta$)
can be analogously accounted for. As a matter of fact, in the $SO(10)$ type of  models
$\text{Tr }Y_{N}\,Y_{N}^{\dagger}\sim \text{Tr }Y_{U}\,Y_{U}^{\dagger}$
due to the common origin of $Y_{U}$ and $Y_{N}$.
The unified structure of the Yukawa sector yields therefore
homogeneous $\Delta a_{i}$ factors
(see the equality of $\sum_{k}y_{ik}$ in \Table{tab:Yukawas}).
This provides the observed large suppression of the Yukawa effects on threshold scales and unification compared to typical two-loop gauge contributions.

In summary, the two-loop RGE effects due to Yukawa couplings on the magnitude of the unification scale (and intermediate thresholds)
and the value of the GUT gauge coupling turn out to be very small. Typically we observe
negative shifts at the per-mil level in both $n_U$
and $\alpha_{U}$, with no relevant impact on the gauge-mediated proton decay rate.

\subsection{The privilege of being minimal}
\label{sec:discussion}

With all the information at hand we can finally approach an assessment of the
viability of the various scenarios.
As we have argued at length, we cannot discard a marginal
unification setup without a detailed information on the fine threshold structure.

Obtaining this piece of information involve the study of the vacuum
of the model, and for $SO(10)$ GUTs this is in general a most challenging task.
In this respect supersymmetry helps: the superpotential is holomorphic and the couplings in the renormalizable case are limited to at most cubic terms;
the physical vacuum is constrained by GUT-scale $F$- and $D$-flatness
and supersymmetry may
be exploited to studying the fermionic rather than the scalar spectra.

It is not surprising that for non-supersymmetric $SO(10)$, only a few detailed studies of the Higgs potential and the related threshold effects
(see for instance
Refs.~\cite{Yasue:1980fy,Harvey:1981hk,Anastaze:1983zk,Babu:1984mz,Abud:1989uu})
are available.
In view of all this and of the intrinsic predictivity related to minimality, the relevance of
carefully scrutinizing the simplest scenarios is hardly overstressed.

The most economical $SO(10)$ Higgs sector includes the adjoint $45_{H}$, that provides the breaking of the GUT symmetry, either $\overline{16}_{H}$ or $\overline{126}_{H}$, responsible for the subsequent $B-L$ breaking, and $10_{H}$, participating to the electroweak symmetry breaking.
The latter is needed together with $\overline{16}_{H}$ or $\overline{126}_{H}$
in order to obtain realistic patterns for the fermionic masses and mixing.
Due to the properties of the adjoint representation this scenario exhibits a minimal
number of parameters in the Higgs potential.
In the current notation such a {\it minimal non-supersymmetric $SO(10)$} (MSO10) GUT
corresponds to the chains VIII and XII.

From this point of view, it is quite intriguing that our analysis of the gauge unification constraints improves the standing of these chains (for XIIa dramatically) with respect to
existing studies.
In particular, considering the renormalizable setups ($\overline{126}_H$), we find
for chain VIIIa, $n_1 \leq 9.1$, $n_U=16.2$ and $\alpha^{-1}_U=45.4$
(to be compared to $n_1 \leq 7.7$ given in Ref.~\cite{Deshpande:1992em}).
This is due to the combination of the updated weak scale data and two loop running effects.
For chain XIIa we find $n_1 \leq 10.8$, $n_U=14.6$ and $\alpha^{-1}_U=44.1$, showing a dramatic (and pathological) change from $n_{1} \leq 5.3$ obtained in \cite{Deshpande:1992em}.
Our result sets the $B-L$ scale nearby the needed scale for realistic light neutrino masses.

We observe non-negligible two-loop effects for the chains VIIIb and XIIb
($\overline{16}_H$) as well.
For chain VIIIb we obtain $n_1 \leq 10.5$, $n_U=16.2$ and $\alpha^{-1}_U=45.6$
(that lifts the $B-L$ scale while preserving $n_{U}$ well above the proton decay bound \eq{pdecaybound}).
A similar shift in $n_{1}$ is observed in chain XIIb where we find
$n_1 \leq 12.5$, $n_U=14.8$ and $\alpha^{-1}_U=44.3$.
As we have already stressed one should not too readily discard $n_U=14.8$ as being incompatible with the proton decay bound. We have verified that reasonable GUT threshold patterns exist that easily lift $n_{U}$ above the experimental bound.
For all these chains D-parity is broken at the GUT scale thus avoiding any cosmological issues
(see the discussion in \sect{sec:chains}).

As remarked in \sect{sec:2loopgauge}, the limit $n_1 = n_2$ leads
to an effective two-step $SO(10)\to \mbox{G2}\to \mbox{SM}$ scenario with a {\em non-minimal} set of surviving scalars at the G2 stage. As a consequence,
the unification setup for the MSO10 can be recovered (with the needed minimal fine tuning) by considering the limit $n_2 = n_U$ in those
chains among I to VII where G1 is either $2_L 2_R 1_X 3_c$ or $2_L 1_R 4_C$
(see \Table{tab:chains}). From inspection of \figs{fig:1to12a}{fig:1to12b} and of \Table{tab:alphaU}, one reads the following results:
for $SO(10)\chain{}{45} 2_L 2_R 1_X 3_c \to \mbox{SM}$
we find
$$
n_1 = 9.5,\ n_U=16.2\ \mbox{and}\ \alpha^{-1}_U=45.5,\ \mbox{in case}\ a
$$
and
$$
n_1 = 10.8,\ n_U=16.2\ \mbox{and}\ \alpha^{-1}_U=45.7,\ \mbox{in case}\ b\ ,
$$
while for $SO(10)\chain{}{45} 2_L 1_R 4_C \to \mbox{SM}$
$$
n_1 = 11.4,\ n_U=14.4\ \mbox{and}\ \alpha^{-1}_U=44.1,\  \mbox{in case}\ a
$$
and
$$
n_1 = 12.6,\ n_U=14.6\ \mbox{and}\ \alpha^{-1}_U=44.3,\ \mbox{in case}\ b\ .
$$
We observe that the patterns are quite similar to those of the non-minimal setups
obtained from chains VIII and XII in the $n_1=n_2$ limit.
Adding the $\phi^{126}$ multiplet , as required by a realistic
matter spectrum in case $a$, does not modify the scalar content in the $2_L 2_R 1_X 3_c$ case:
only one linear combination of the $10_H$ and $\overline{126}_H$ bidoublets (see \Table{tab:submultiplets})
is allowed by minimal fine tuning. On the other hand, in the $2_L 1_R 4_C$ case,
the only sizeable effect is a shift on the unified coupling constant, namely $\alpha^{-1}_U=40.7$
(see the discussion in \sect{sec:extrahiggs}).

In summary, in view of realistic thresholds effects at the GUT (and $B-L$) scale and of
a modest fine tuning in the see-saw neutrino mass, we consider both scenarios
worth of a detailed investigation.

\section{Outlook}
\label{sec:outlook}

We presented an updated and systematic two-loop discussion of
non-supersymmetric $SO(10)$ gauge unification with two (and one) intermediate scales.
We completed and corrected existing analyses by including a thorough discussion
of $U(1)$ mixing, which affects the gauge running already at the one-loop level
in a number of interesting $SO(10)$ breaking chains. We assessed the relevance
of additional Higgs multiplets, needed at some of the intermediate stages
in order to reproduce a realistic fermionic mass spectrum.
Finally, we found and fixed several discrepancies in the two-loop $\beta$-coefficients.

The updated results have a non-negligible impact on the
values of the unification and $B-L$ scales (as well as on the value of the unified gauge coupling).
This is due to the combined effects of the one-loop dynamics corresponding to the $U(1)$
gauge mixing and of the two-loop RGE contributions.

We discussed the viability of the different $SO(10)$ scenarios on the basis of
proton decay and the see-saw induced neutrino mass. We were lead to focus our attention
on the minimal $SO(10)$ setup, emerging from a balance of minimal dimensionality
Higgs representations and a minimal number of parameters in the scalar potential.
Such a scenario invokes, in addition to a complex $10_H$, one adjoint $45_H$
together with one $\overline{126}_H$ or $\overline{16}_H$ at the effective level.

Although the updated values of the unification or $B-L$ scales
are in some of the setups still conflicting with the experimental requirements,
they are close enough that reasonable spreads
in the GUT thresholds (or a moderate fine tuning in the neutrino mass
matrix) can easily restore the agreement.
This may entail the detailed study of the scalar potential of the model beyond the tree
approximation, that is a rather non-trivial task.
Nevertheless, the appeal of minimality (with supersymmetry
confined to the Planck scale) motivates us to pursue this study.


\subsection*{Acknowledgments}

S.B. acknowledges support by MIUR and by the RTN European Program
MRTN-CT-2004-503369. The work of M.M. is supported by the Royal Institute of
Technology (KTH), Contract No. SII-56510. M.M. is grateful to SISSA
for the hospitality during the preparation of part of the manuscript.


\appendix

\section{One- and Two-loop beta coefficients}
\label{app:2Lbeta}

In this appendix we report the one- and two-loop $\beta$-coefficients
used in the numerical analysis. The calculation of the $U(1)$ mixing coefficients
and of the Yukawa contributions to the gauge coupling renormalization is detailed in Apps. \ref{app:U1mix} and
\ref{app:Yukawa} respectively.

\begin{table*}
\begin{tabular}{lcclcc}
\hline
\multicolumn{6}{c}{G2 ($M_{U}$ $\rightarrow$ $M_{2}$)} \\
\hline
{\rm Chain}  &  {\rm $a_j$}  &  {\rm $b_{ij}$}  & {\rm Chain}  &  {\rm $a_j$}  &  {\rm $b_{ij}$} \\
\hline

\sepB
{\rm  Ia}  &
$(-3, \frac{11}{3}, -7)$ &
$\left(
\begin{array}{ccc}
 8 & 3 & \frac{45}{2} \\
 3 & \frac{584}{3} & \frac{765}{2} \\
 \frac{9}{2} & \frac{153}{2} & \frac{289}{2}
\end{array}
\right)$
&
{\rm  Ib}  &
$(-3, -\frac{7}{3}, -\frac{29}{3})$ &
$\left(
\begin{array}{ccc}
 8 & 3 & \frac{45}{2} \\
 3 & \frac{50}{3} & \frac{75}{2} \\
 \frac{9}{2} & \frac{15}{2} & -\frac{94}{3}
\end{array}
\right)$
\\

{\rm  IIa}  &
$({\frac{11}{3}, \frac{11}{3}, -4})$ &
$\left(
\begin{array}{ccc}
 \frac{584}{3} & 3 & \frac{765}{2} \\
 3 & \frac{584}{3} & \frac{765}{2} \\
 \frac{153}{2} & \frac{153}{2} & \frac{661}{2}
\end{array}
\right)$
&
{\rm  IIb}  &
$({-\frac{7}{3}, -\frac{7}{3}, -\frac{28}{3}})$ &
$\left(
\begin{array}{ccc}
 \frac{50}{3} & 3 & \frac{75}{2} \\
 3 & \frac{50}{3} & \frac{75}{2} \\
 \frac{15}{2} & \frac{15}{2} & -\frac{127}{6}
\end{array}
\right)$
\\

\sepB
{\rm  IIIa}  &
$({\frac{11}{3}, \frac{11}{3}, -4})$ &
$\left(
\begin{array}{ccc}
 \frac{584}{3} & 3 & \frac{765}{2} \\
 3 & \frac{584}{3} & \frac{765}{2} \\
 \frac{153}{2} & \frac{153}{2} & \frac{661}{2}
\end{array}
\right)$
&
{\rm  IIIb}  &
$({-\frac{7}{3}, -\frac{7}{3}, -\frac{28}{3}})$ &
$\left(
\begin{array}{ccc}
 \frac{50}{3} & 3 & \frac{75}{2} \\
 3 & \frac{50}{3} & \frac{75}{2} \\
 \frac{15}{2} & \frac{15}{2} & -\frac{127}{6}
\end{array}
\right)$
\\

\sepB
{\rm IVa}  &
$({-\frac{7}{3}, -\frac{7}{3}, 7, -7})$ &
$\left(
\begin{array}{cccc}
 \frac{80}{3} & 3 & \frac{27}{2} & 12 \\
 3 & \frac{80}{3} & \frac{27}{2} & 12 \\
 \frac{81}{2} & \frac{81}{2} & \frac{115}{2} & 4 \\
 \frac{9}{2} & \frac{9}{2} & \frac{1}{2} & -26
\end{array}
\right)$
&
{\rm  IVb}  &
$(-\frac{17}{6}, -\frac{17}{6}, \frac{9}{2}, -7)$ &
$\left(
\begin{array}{cccc}
 \frac{61}{6} & 3 & \frac{9}{4} & 12 \\
 3 & \frac{61}{6} & \frac{9}{4} & 12 \\
 \frac{27}{4} & \frac{27}{4} & \frac{23}{4} & 4 \\
 \frac{9}{2} & \frac{9}{2} & \frac{1}{2} & -26
\end{array}
\right)$
\\

\sepA
{\rm  Va}  &
$(-3, 4, -\frac{23}{3})$ &
$\left(
\begin{array}{ccc}
 8 & 3 & \frac{45}{2} \\
 3 & 204 & \frac{765}{2} \\
 \frac{9}{2} & \frac{153}{2} & \frac{643}{6}
\end{array}
\right)$
&
{\rm  Vb}  &
$(-3, -2, -\frac{31}{3})$ &
$\left(
\begin{array}{ccc}
 8 & 3 & \frac{45}{2} \\
 3 & 26 & \frac{75}{2} \\
 \frac{9}{2} & \frac{15}{2} & -\frac{206}{3}
\end{array}
\right)$
\\

\sepA
{\rm  VIa}  &
$(4, 4, -\frac{14}{3})$ &
$\left(
\begin{array}{ccc}
 204 & 3 & \frac{765}{2} \\
 3 & 204 & \frac{765}{2} \\
 \frac{153}{2} & \frac{153}{2} & \frac{1759}{6}
\end{array}
\right)$
&
{\rm  VIb}  &
$(-2, -2, -10)$ &
$\left(
\begin{array}{ccc}
 26 & 3 & \frac{75}{2} \\
 3 & 26 & \frac{75}{2} \\
 \frac{15}{2} & \frac{15}{2} & -\frac{117}{2}
\end{array}
\right)$
\\

\sepA
{\rm  VIIa}  &
$(\frac{11}{3}, \frac{11}{3}, -\frac{14}{3})$ &
$\left(
\begin{array}{ccc}
 \frac{584}{3} & 3 & \frac{765}{2} \\
 3 & \frac{584}{3} & \frac{765}{2} \\
 \frac{153}{2} & \frac{153}{2} & \frac{1759}{6}
\end{array}
\right)$
&
{\rm  VIIb}  &
$(-\frac{7}{3}, -\frac{7}{3}, -10)$ &
$\left(
\begin{array}{ccc}
 \frac{50}{3} & 3 & \frac{75}{2} \\
 3 & \frac{50}{3} & \frac{75}{2} \\
 \frac{15}{2} & \frac{15}{2} & -\frac{117}{2}
\end{array}
\right)$
\\

\sepB
{\rm  VIIIa}  &
$(-3, -2, \frac{11}{2}, -7)$ &
$\left(
\begin{array}{cccc}
 8 & 3 & \frac{3}{2} & 12 \\
 3 & 36 & \frac{27}{2} & 12 \\
 \frac{9}{2} & \frac{81}{2} & \frac{61}{2} & 4 \\
 \frac{9}{2} & \frac{9}{2} & \frac{1}{2} & -26
\end{array}
\right)$
&
{\rm  VIIIb}  &
$(-3, -\frac{5}{2}, \frac{17}{4}, -7)$ &
$\left(
\begin{array}{cccc}
 8 & 3 & \frac{3}{2} & 12 \\
 3 & \frac{39}{2} & \frac{9}{4} & 12 \\
 \frac{9}{2} & \frac{27}{4} & \frac{37}{8} & 4 \\
 \frac{9}{2} & \frac{9}{2} & \frac{1}{2} & -26
\end{array}
\right)$
\\

\sepB
{\rm  IXa}  &
$(-2, -2, 7, -7)$ &
$\left(
\begin{array}{cccc}
 36 & 3 & \frac{27}{2} & 12 \\
 3 & 36 & \frac{27}{2} & 12 \\
 \frac{81}{2} & \frac{81}{2} & \frac{115}{2} & 4 \\
 \frac{9}{2} & \frac{9}{2} & \frac{1}{2} & -26
\end{array}
\right)$
&
{\rm  IXb}  &
$(-\frac{5}{2}, -\frac{5}{2}, \frac{9}{2}, -7)$ &
$\left(
\begin{array}{cccc}
 \frac{39}{2} & 3 & \frac{9}{4} & 12 \\
 3 & \frac{39}{2} & \frac{9}{4} & 12 \\
 \frac{27}{4} & \frac{27}{4} & \frac{23}{4} & 4 \\
 \frac{9}{2} & \frac{9}{2} & \frac{1}{2} & -26
\end{array}
\right)$
\\

\sepB
{\rm  Xa}  &
$(-3, \frac{26}{3}, -\frac{17}{3})$ &
$\left(
\begin{array}{ccc}
 8 & 3 & \frac{45}{2} \\
 3 & \frac{1004}{3} & \frac{1245}{2} \\
 \frac{9}{2} & \frac{249}{2} & \frac{1315}{6}
\end{array}
\right)$
&
{\rm  Xb}  &
$(-3, \frac{8}{3}, -\frac{25}{3})$ &
$\left(
\begin{array}{ccc}
 8 & 3 & \frac{45}{2} \\
 3 & \frac{470}{3} & \frac{555}{2} \\
 \frac{9}{2} & \frac{111}{2} & \frac{130}{3}
\end{array}
\right)$
\\

{\rm  XIa}  &
$(\frac{26}{3}, \frac{26}{3}, -\frac{2}{3})$ &
$\left(
\begin{array}{ccc}
 \frac{1004}{3} & 3 & \frac{1245}{2} \\
 3 & \frac{1004}{3} & \frac{1245}{2} \\
 \frac{249}{2} & \frac{249}{2} & \frac{3103}{6}
\end{array}
\right)$
&
{\rm  XIb}  &
$(\frac{8}{3}, \frac{8}{3}, -6)$ &
$\left(
\begin{array}{ccc}
 \frac{470}{3} & 3 & \frac{555}{2} \\
 3 & \frac{470}{3} & \frac{555}{2} \\
 \frac{111}{2} & \frac{111}{2} & \frac{331}{2}
\end{array}
\right)$
\\

\sepB
{\rm  XIIa}  &
$(-\frac{19}{6}, \frac{15}{2}, -9)$ &
$\left(
\begin{array}{ccc}
 \frac{35}{6} & \frac{1}{2} & \frac{45}{2} \\
 \frac{3}{2} & \frac{87}{2} & \frac{405}{2} \\
 \frac{9}{2} & \frac{27}{2} & \frac{41}{2}
\end{array}
\right)$
&
{\rm  XIIb}  &
$(-\frac{19}{6}, \frac{9}{2}, -\frac{59}{6})$ &
$\left(
\begin{array}{ccc}
 \frac{35}{6} & \frac{1}{2} & \frac{45}{2} \\
 \frac{3}{2} & \frac{9}{2} & 30 \\
 \frac{9}{2} & 2 & -\frac{437}{12}
\end{array}
\right)$
\\

\hline
\end{tabular}
\mycaption{The $a_{i}$ and $b_{ij}$ coefficients due to pure gauge interactions are
reported for
the G2 chains with $\overline{126}_{H}$ (left) and ${16}_{H}$ (right) respectively.
The two-loop contributions induced by Yukawa couplings are given
in Appendix \ref{app:Yukawa}}
\label{tab:beta-G2}
\end{table*}


\begin{table*}
\begin{tabular}{cccccc}
\hline
\multicolumn{6}{c}{G1 ($M_{2} \rightarrow M_{1}$)} \\
\hline
{\rm Chain}  &  {\rm $a_i$}  &  {\rm $b_{ij}$}  & {\rm Chain}  &  {\rm $a_i$}  &  {\rm $b_{ij}$} \\
\hline
\\[-2ex]
\sepB
{\rm Ia}  &
$(-3, -\frac{7}{3}, \frac{11}{2}, -7)$ &
$\left(
\begin{array}{cccc}
 8 & 3 & \frac{3}{2} & 12 \\
 3 & \frac{80}{3} & \frac{27}{2} & 12 \\
 \frac{9}{2} & \frac{81}{2} & \frac{61}{2} & 4 \\
 \frac{9}{2} & \frac{9}{2} & \frac{1}{2} & -26
\end{array}
\right)$
&
{\rm  Ib}  &
$(-3, -\frac{17}{6}, \frac{17}{4}, -7)$ &
$\left(
\begin{array}{cccc}
 8 & 3 & \frac{3}{2} & 12 \\
 3 & \frac{61}{6} & \frac{9}{4} & 12 \\
 \frac{9}{2} & \frac{27}{4} & \frac{37}{8} & 4 \\
 \frac{9}{2} & \frac{9}{2} & \frac{1}{2} & -26
\end{array}
\right)$
\\

\sepB
{\rm  IIa}  &
$(-\frac{7}{3}, -\frac{7}{3}, 7, -7)$ &
$\left(
\begin{array}{cccc}
 \frac{80}{3} & 3 & \frac{27}{2} & 12 \\
 3 & \frac{80}{3} & \frac{27}{2} & 12 \\
 \frac{81}{2} & \frac{81}{2} & \frac{115}{2} & 4 \\
 \frac{9}{2} & \frac{9}{2} & \frac{1}{2} & -26
\end{array}
\right)$
&
{\rm  IIb}  &
$(-\frac{17}{6}, -\frac{17}{6}, \frac{9}{2}, -7)$ &
$\left(
\begin{array}{cccc}
 \frac{61}{6} & 3 & \frac{9}{4} & 12 \\
 3 & \frac{61}{6} & \frac{9}{4} & 12 \\
 \frac{27}{4} & \frac{27}{4} & \frac{23}{4} & 4 \\
 \frac{9}{2} & \frac{9}{2} & \frac{1}{2} & -26
\end{array}
\right)$
\\

\sepB
{\rm IIIa}  &
$(-3, -\frac{7}{3}, \frac{11}{2}, -7)$ &
$\left(
\begin{array}{cccc}
 8 & 3 & \frac{3}{2} & 12 \\
 3 & \frac{80}{3} & \frac{27}{2} & 12 \\
 \frac{9}{2} & \frac{81}{2} & \frac{61}{2} & 4 \\
 \frac{9}{2} & \frac{9}{2} & \frac{1}{2} & -26
\end{array}
\right)$
&
{\rm  IIIb}  &
$(-3, -\frac{17}{6}, \frac{17}{4}, -7)$ &
$\left(
\begin{array}{cccc}
 8 & 3 & \frac{3}{2} & 12 \\
 3 & \frac{61}{6} & \frac{9}{4} & 12 \\
 \frac{9}{2} & \frac{27}{4} & \frac{37}{8} & 4 \\
 \frac{9}{2} & \frac{9}{2} & \frac{1}{2} & -26
\end{array}
\right)$
\\

\sepB
{\rm IVa}  &
$(-3, -\frac{7}{3}, \frac{11}{2}, -7)$ &
$\left(
\begin{array}{cccc}
 8 & 3 & \frac{3}{2} & 12 \\
 3 & \frac{80}{3} & \frac{27}{2} & 12 \\
 \frac{9}{2} & \frac{81}{2} & \frac{61}{2} & 4 \\
 \frac{9}{2} & \frac{9}{2} & \frac{1}{2} & -26
\end{array}
\right)$
 &
{\rm   IVb}  &
$(-3, -\frac{17}{6}, \frac{17}{4}, -7)$ &
$\left(
\begin{array}{cccc}
 8 & 3 & \frac{3}{2} & 12 \\
 3 & \frac{61}{6} & \frac{9}{4} & 12 \\
 \frac{9}{2} & \frac{27}{4} & \frac{37}{8} & 4 \\
 \frac{9}{2} & \frac{9}{2} & \frac{1}{2} & -26
\end{array}
\right)$
\\

\sepA
{\rm  Va}  &
$(-\frac{19}{6}, \frac{15}{2}, -\frac{29}{3})$ &
$\left(
\begin{array}{ccc}
 \frac{35}{6} & \frac{1}{2} & \frac{45}{2} \\
 \frac{3}{2} & \frac{87}{2} & \frac{405}{2} \\
 \frac{9}{2} & \frac{27}{2} & -\frac{101}{6}
\end{array}
\right)$
&
{\rm  Vb}  &
$(-\frac{19}{6}, \frac{9}{2}, -\frac{21}{2})$ &
$\left(
\begin{array}{ccc}
 \frac{35}{6} & \frac{1}{2} & \frac{45}{2} \\
 \frac{3}{2} & \frac{9}{2} & 30 \\
 \frac{9}{2} & 2 & -\frac{295}{4}
\end{array}
\right)$
\\

\sepA
{\rm  VIa}  &
$(-\frac{19}{6}, \frac{15}{2}, -\frac{29}{3})$ &
$\left(
\begin{array}{ccc}
 \frac{35}{6} & \frac{1}{2} & \frac{45}{2} \\
 \frac{3}{2} & \frac{87}{2} & \frac{405}{2} \\
 \frac{9}{2} & \frac{27}{2} & -\frac{101}{6}
\end{array}
\right)$
&
{\rm  VIb}  &
$(-\frac{19}{6}, \frac{9}{2}, -\frac{21}{2})$ &
$\left(
\begin{array}{ccc}
 \frac{35}{6} & \frac{1}{2} & \frac{45}{2} \\
 \frac{3}{2} & \frac{9}{2} & 30 \\
 \frac{9}{2} & 2 & -\frac{295}{4}
\end{array}
\right)$
\\

\sepA
{\rm  VIIa}  &
$(-3, \frac{11}{3}, -\frac{23}{3})$ &
$\left(
\begin{array}{ccc}
 8 & 3 & \frac{45}{2} \\
 3 & \frac{584}{3} & \frac{765}{2} \\
 \frac{9}{2} & \frac{153}{2} & \frac{643}{6}
\end{array}
\right)$
&
{\rm  VIIb}  &
$(-3, -\frac{7}{3}, -\frac{31}{3})$ &
$\left(
\begin{array}{ccc}
 8 & 3 & \frac{45}{2} \\
 3 & \frac{50}{3} & \frac{75}{2} \\
 \frac{9}{2} & \frac{15}{2} & -\frac{206}{3}
\end{array}
\right)$
\\

\hline
\end{tabular}
\mycaption{The $a_{i}$ and $b_{ij}$ coefficients due to gauge interactions are reported for the G1 chains I to VII with $\overline{126}_{H}$ (left) and $\overline{16}_{H}$ (right) respectively.
The two-loop contributions induced by Yukawa couplings are given
in Appendix~\ref{app:Yukawa}}
\label{tab:beta-G1}
\end{table*}

\begin{table*}
\begin{tabular}{cccccc}
\hline
\multicolumn{6}{c}{G1 ($M_{2} \rightarrow M_{1}$)} \\
\hline
{\rm Chain}  &  {\rm $a_i$}  &  {\rm $b_{ij}$}  & {\rm Chain}  &  {\rm $a_i$}  &  {\rm $b_{ij}$} \\
\hline

\sepC
$\begin{array}{c}
{\rm  VIIIa} \\
\vdots \\
{\rm  XIIa}
\end{array}$
 &
\;\;$(-\frac{19}{6}, \frac{9}{2}, \frac{9}{2}, -7)$ &
$\left(
\begin{array}{cccc}
 \frac{35}{6} & \frac{1}{2} & \frac{3}{2} & 12 \\
 \frac{3}{2} & \frac{15}{2} & \frac{15}{2} & 12 \\
 \frac{9}{2} & \frac{15}{2} & \frac{25}{2} & 4 \\
 \frac{9}{2} & \frac{3}{2} & \frac{1}{2} & -26
\end{array}
\right)$\;\;
&
\sepB
$\begin{array}{c}
{\rm  VIIIb} \\
\vdots \\
{\rm  XIIb}
\end{array}$
&
\;\;$(-\frac{19}{6},  \frac{17}{4}, \frac{33}{8}, -7)$&
$\left(
\begin{array}{cccc}
 \frac{35}{6} & \frac{1}{2} & \frac{3}{2} & 12 \\
 \frac{3}{2} & \frac{15}{4} & \frac{15}{8} & 12 \\
 \frac{9}{2} & \frac{15}{8} & \frac{65}{16} & 4 \\
 \frac{9}{2} & \frac{3}{2} & \frac{1}{2} & -26
\end{array}
\right)$
\\

\hline
\end{tabular}
\mycaption{The $a_{i}$ and $b_{ij}$ coefficients due to purely gauge interactions for the
G1 chains VIII to XII are reported.
For comparison with previous studies the $\beta$-coefficients
are given neglecting systematically one- and two-loops $U(1)$ mixing effects
(while all diagonal $U(1)$ contributions
to abelian and non-abelian gauge coupling renormalization are included).
The complete (and correct) treatment of $U(1)$ mixing
is detailed in Appendix~\ref{app:U1mix}.}
\label{tab:beta-U1nomix}
\end{table*}

\begin{table}
\begin{tabular}{ccc}
\hline
\multicolumn{3}{c}{SM ($M_{1} \rightarrow M_Z$)} \\
\hline
{\rm Chain}  &  {\rm $a_i$}  &  {\rm $b_{ij}$}
\\
\hline
\sepB
{\rm All}  &
$(\frac{41}{10}, -\frac{19}{6}, -7)$ &
$\left(
\begin{array}{ccc}
 \frac{199}{50} & \frac{27}{10} & \frac{44}{5} \\
 \frac{9}{10} & \frac{35}{6} & 12 \\
 \frac{11}{10} & \frac{9}{2} & -26
\end{array}
\right)$
\\
\hline
\end{tabular}
\mycaption{The $a_i$ and $b_{ij}$ coefficients are given for the $1_Y 2_L 3_c$ (SM) gauge running. The scalar sector includes one Higgs doublet.}
\label{tab:beta-SM}
\end{table}

\begin{table}
\begin{tabular}{lcc}
\hline
{\rm Chain}
&
$\tilde{b}_{ij}$
&
{\rm Eq. in Ref. \cite{Chang:1984qr}\ }
\\
\hline
\\
{\rm All/SM}
&
$\left(
\begin{array}{ccc}
 \frac{199}{205} & -\frac{81}{95} & -\frac{44}{35} \\
 \frac{9}{41} & -\frac{35}{19} & -\frac{12}{7} \\
 \frac{11}{41} & -\frac{27}{19} & \frac{26}{7}
\end{array}
\right)$
&
{\rm A7}
\\
\\
{\rm VIIIa/G1}
&
$\left(
\begin{array}{cccc}
 \frac{25}{9} & \frac{5}{3} & -\frac{27}{19} & -\frac{4}{7} \\
 \frac{5}{3} & \frac{5}{3} & -\frac{9}{19} & -\frac{12}{7} \\
 \frac{1}{3} & \frac{1}{9} & -\frac{35}{19} & -\frac{12}{7} \\
 \frac{1}{9} & \frac{1}{3} & -\frac{27}{19} & \frac{26}{7}
\end{array}
\right)$
&
{\rm A10}
\\
\\
{\rm VIIIa/G2}
&
$\left(
\begin{array}{cccc}
 \frac{61}{11} & -\frac{3}{2} & -\frac{81}{4} & -\frac{4}{7} \\
 \frac{3}{11} & -\frac{8}{3} & -\frac{3}{2} & -\frac{12}{7} \\
 \frac{27}{11} & -1 & -18 & -\frac{12}{7} \\
 \frac{1}{11} & -\frac{3}{2} & -\frac{9}{4} & \frac{26}{7}
\end{array}
\right)$
&
{\rm A13}
\\
\\
{\rm Ia/G2}
&
$\left(
\begin{array}{ccc}
 -\frac{8}{3} & \frac{9}{11} & -\frac{45}{14} \\
 -1 & \frac{584}{11} & -\frac{765}{14} \\
 -\frac{3}{2} & \frac{459}{22} & -\frac{289}{14}
\end{array}
\right)$
&
{\rm A14}
\\
\\
{\rm Va/G1}
&
$\left(
\begin{array}{ccc}
 -\frac{35}{19} & \frac{1}{15} & -\frac{135}{58} \\
 -\frac{9}{19} & \frac{29}{5} & -\frac{1215}{58} \\
 -\frac{27}{19} & \frac{9}{5} & \frac{101}{58}
\end{array}
\right)$
&
{\rm A15}
\\
\\
{\rm XIIa/G2}
&
$\left(
\begin{array}{ccc}
 -\frac{35}{19} & \frac{1}{15} & -\frac{5}{2} \\
 -\frac{9}{19} & \frac{29}{5} & -\frac{45}{2} \\
 -\frac{27}{19} & \frac{9}{5} & -\frac{41}{18}
\end{array}
\right)$
&
{\rm A18}
\\
\\
\hline
\end{tabular}
\mycaption{The rescaled two-loop $\beta$-coefficients $\tilde{b}_{ij}$
computed in this paper are shown together with the corresponding equations in Ref.~\cite{Chang:1984qr}.
For the purpose of comparison
Yukawa contributions are neglected and no $U(1)$ mixing is included in chain VIIIa/G1.
Care must be taken of the different ordering between abelian and non-abelian gauge
group factors in Ref.~\cite{Chang:1984qr}.
We report those cases where disagreement is found in some of the entries, while
we fully agree with the Eqs. A9, A11 and A16.
}
\label{tab:betaChang}
\end{table}

\begin{table}
\begin{tabular}{ccc}
\hline
{$\phi^{126}$}  &  {\rm $a_i$}  &  {\rm $b_{ij}$}
\\
\hline
\\
{\rm $(2,2,15)$}  &
$(5, 5, \frac{16}{3})$ &
$\left(
\begin{array}{ccc}
 65 & 45 & 240 \\
 45 & 65 & 240 \\
 48 & 48 & \frac{896}{3}
\end{array}
\right)$
\\
\\
{\rm $(2,+\frac{1}{2},15)$}  &
$(\frac{5}{2}, \frac{5}{2}, \frac{8}{3})$ &
$\left(
\begin{array}{ccc}
 \frac{65}{2} & \frac{15}{2} & 120 \\
 \frac{45}{2} & \frac{15}{2} & 120 \\
 24 & 8 & \frac{448}{3}
\end{array}
\right)$
\\
\\
\hline
\end{tabular}
\caption{One- and two-loop additional contributions to the $\beta$-coefficients
related to the presence of the $\phi^{126}$ scalar multiplets
in the $2_L 2_R 4$ (top) and $2_L 1_R 4$ (bottom) stages.}
\label{tab:phi126betas}
\end{table}

\clearpage

\subsection{Beta-functions with $U(1)$ mixing}
\label{app:U1mix}

The basic building blocks of the one- and two-loop $\beta$-functions for the abelian couplings
with $U(1)$ mixing, c.f. \eqs{bU12loops}{bpU12loops}, can be conveniently
written as
\beq
g_{ka}g_{kb}=g_{sa}\Gamma^{(1)}_{sr}g_{rb}
\label{1loopterm}
\eeq
and
\beq
g_{ka}g_{kb}g^2_{kc}=g_{sa}\Gamma^{(2)}_{sr}g_{rb}
\label{2loopterm}\,,
\eeq
where $\Gamma^{(1)}$ and $\Gamma^{(2)}$ are functions of the abelian charges $Q_{k}^{a}$
and, at two loops, also of the gauge couplings.
In the case of interest, i.e. for two abelian charges $U(1)_A$ and $U(1)_B$,
one obtains
\begin{align}
\Gamma^{(1)}_{AA}&=(Q_k^A)^2\,,
\nn \\[1ex]
\Gamma^{(1)}_{AB}&=\Gamma^{(1)}_{BA}=Q_k^AQ_k^B\,,
\label{1loopgammatilde}
\\[1ex]
\Gamma^{(1)}_{BB}&=(Q_k^B)^2\,, \nn
\end{align}
and
\begin{widetext}
\begin{align}
\Gamma^{(2)}_{AA}&=(Q_k^A)^4(g^2_{AA}+g^2_{AB})+2(Q_k^A)^3Q_k^B(g_{AA}g_{BA}+g_{AB}g_{BB})+
(Q_k^A)^2(Q_k^B)^2(g^2_{BA}+g^2_{BB})\,,
\nn \\[2ex]
\Gamma^{(2)}_{AB}&=\Gamma^{(2)}_{BA}=(Q_k^A)^3Q_k^B(g^2_{AA}+g^2_{AB})+2(Q_k^A)^2(Q_k^B)^2(g_{AA}g_{BA}+g_{AB}g_{BB})+
Q_k^A(Q_k^B)^3(g^2_{BA}+g^2_{BB})\,,
\label{2loopgammatilde}
\\[2ex]
\Gamma^{(2)}_{BB}&=(Q_k^A)^2(Q_k^B)^2(g^2_{AA}+g^2_{AB})+2Q_k^A(Q_k^B)^3(g_{AA}g_{BA}+g_{AB}g_{BB})+
(Q_k^B)^4(g^2_{BA}+g^2_{BB}) \,.\nn
\end{align}
\end{widetext}

All other contributions
in \eq{bU12loops} and \eq{bpU12loops} can be easily obtained
from \eqs{1loopgammatilde}{2loopgammatilde} by including the appropriate group factors.
It is worth mentioning that for complete $SO(10)$ multiplets,
$(Q_k^A)^n(Q_k^B)^m = 0$ for n and m odd
(with $n+m=2$ at one-loop and $n+m=4$ at two-loop level).

By evaluating \eqs{1loopgammatilde}{2loopgammatilde} for the particle content relevant to
the $2_L1_R1_X3_c$ stages in chains VIII-XII, and by substituting into \eqs{bU12loops}{bpU12loops},
one finally obtains

\begin{widetext}
\begin{itemize}

\item Chains VIII-XII with $\overline{126}_{H}$ in the Higgs sector:

\begin{align}
\gamma_{RR} &= \frac{9}{2}+\frac{1}{(4\pi)^2}\left[\frac{15}{2}(g_{RR}^2+g_{RX}^2)-4\sqrt{6}(g_{RR}g_{XR}+g_{RX}g_{XX})+
\frac{15}{2}(g_{XR}^2+g_{XX}^2)+\frac{3}{2}g^2_L+12g^2_c\right]\,, \nn \\[1ex]
\gamma_{RX} &= \gamma_{XR}=-\frac{1}{\sqrt{6}}+\frac{1}{(4\pi)^2}\left[-2{\sqrt{6}}(g_{RR}^2+g_{RX}^2)+
15(g_{RR}g_{XR}+g_{RX}g_{XX})-3\sqrt{6}(g_{XR}^2+g_{XX}^2)\right]\,, \nn \\[1ex]
\gamma_{XX} &= \frac{9}{2}+\frac{1}{(4\pi)^2}\left[\frac{15}{2}(g_{RR}^2+g_{RX}^2)-6\sqrt{6}(g_{RR}g_{XR}+g_{RX}g_{XX})+
\frac{25}{2}(g_{XR}^2+g_{XX}^2)+\frac{9}{2}g^2_L+4g^2_c\right]
\label{gammaXR_1}\,, \\[1ex]
\gamma_L &= -\frac{19}{6}+\frac{1}{(4\pi)^2}\left[\frac{1}{2}(g_{RR}^2+g_{RX}^2)+
\frac{3}{2}(g_{XR}^2+g_{XX}^2)+\frac{35}{6}g^2_L+12g^2_c\right] \,,\nn \\[1ex]
\gamma_c &= -7+\frac{1}{(4\pi)^2}\left[\frac{3}{2}(g_{RR}^2+g_{RX}^2)+
\frac{1}{2}(g_{XR}^2+g_{XX}^2)+\frac{9}{2}g^2_L-26g^2_c\right]\,; \nn
\end{align}

\item Chains VIII-XII with $\overline{16}_{H}$ in the Higgs sector:

\begin{align}
\gamma_{RR} &= \frac{17}{4}+\frac{1}{(4\pi)^2}\left[\frac{15}{4}(g_{RR}^2+g_{RX}^2)-\frac{1}{2}\sqrt{\frac{3}{2}}(g_{RR}g_{XR}+g_{RX}g_{XX})+
\frac{15}{8}(g_{XR}^2+g_{XX}^2)+\frac{3}{2}g^2_L+12g^2_c\right]\,, \nn \\[1ex]
\gamma_{RX} &= \gamma_{XR}=-\frac{1}{4\sqrt{6}}+\frac{1}{(4\pi)^2}\left[-\frac{1}{4}\sqrt{\frac{3}{2}}(g_{RR}^2+g_{RX}^2)+
\frac{15}{4}(g_{RR}g_{XR}+g_{RX}g_{XX})-\frac{3}{8}\sqrt{\frac{3}{2}}(g_{XR}^2+g_{XX}^2)\right]\,, \nn \\[1ex]
\gamma_{XX} &= \frac{33}{8}+\frac{1}{(4\pi)^2}\left[\frac{15}{8}(g_{RR}^2+g_{RX}^2)-\frac{3}{4}\sqrt{\frac{3}{2}}(g_{RR}g_{XR}+g_{RX}g_{XX})
+\frac{65}{16}(g_{XR}^2+g_{XX}^2)+\frac{9}{2}g^2_L+4g^2_c\right]
\label{gammaXR_2}\,, \nn \\[1ex]
\gamma_L &= -\frac{19}{6}+\frac{1}{(4\pi)^2}\left[\frac{1}{2}(g_{RR}^2+g_{RX}^2)+
\frac{3}{2}(g_{XR}^2+g_{XX}^2)+\frac{35}{6}g^2_L+12g^2_c\right]\,,  \\[1ex]
\gamma_c &= -7+\frac{1}{(4\pi)^2}\left[\frac{3}{2}(g_{RR}^2+g_{RX}^2)+
\frac{1}{2}(g_{XR}^2+g_{XX}^2)+\frac{9}{2}g^2_L-26g^2_c\right]\,. \nn
\end{align}

\end{itemize}
\end{widetext}

By setting  $\gamma_{XR}=\gamma_{RX}=0$ and $g_{XR}=g_{RX}=0$ in \eqs{gammaXR_1}{gammaXR_2}
one obtains the one- and two-loop $\beta$-coefficients in the diagonal approximation, as reported
in \Table{tab:beta-U1nomix}. The latter are used in \figs{fig:1to12a}{fig:1to12b}
for the only purpose of exhibiting the effect of the abelian mixing in the gauge coupling
renormalization.

\subsection{Yukawa contributions}
\label{app:Yukawa}
The Yukawa couplings enter the gauge $\beta$-functions first at the two-loop level,
c.f. \eq{Gp2loops} and \eq{bU12loops}. Since the notation adopted in
\eqs{Y4}{Yukawa} is rather concise we shall detail the structure
of \eq{Y4}, paying particular attention to the calculation of the $y_{pk}$
coefficients in \eq{Ycpk}.

The trace on the RHS of \eq{Y4} is taken over all indices of the fields
entering the Yukawa interaction in \eq{Yukawa}. Considering for instance the up-quark
Yukawa sector of the SM the term
$ \overline{Q_{L}}Y_{U}U_{R}\tilde{h}+h.c.$
(with $\tilde h=i\sigma_{2}h^{*}$) can be explicitly written as
\begin{equation}
\label{Yukawaguts}
Y^{ab}_{U}{\varepsilon}^{kl}{\delta_{3}}_{j}^{i}\overline{Q_{L}^{a}}_{ik}U^{bj}_{R}h^{*}_{l}+h.c.\,,
\end{equation}
where $\{a,b\}$, $\{i,j\}$ and $\{k,l\}$ label flavour, $SU(3)_{c}$ and
$SU(2)_{L}$ indices respectively, while $\delta_{n}$ denotes the $n$-dimensional Kronecker $\delta$ symbol.
Thus, the Yukawa coupling entering \eq{Y4} is a 6-dimensional
object with the index structure
$Y^{ab}_{U}{\varepsilon}^{kl}{\delta_{3}}_{j}^{i}$.
The contribution of \eq{Yukawaguts} to the three $y_{pU}$ coefficients
(conveniently separated into two terms corresponding to the fermionic representations
$Q_{L}$ and $U_{R}$) can then be written as
\begin{eqnarray}
y_{pU}&=& \frac{1}{d(G_{p})}
\left[C_{2}^{(p)}(Q_{L})+C_{2}^{(p)}(U_{R})\right] \nn \\
&&
\times \sum_{ab,ij,kl}Y^{ab}_{U}{\varepsilon}^{kl}{\delta_{3}}_{j}^{i}Y^{ab*}_{U}
{\varepsilon}_{kl}{\delta_{3}}_{i}^{j}
\label{ypU}
\end{eqnarray}
The sum can be factorized into the flavour space part
$\sum_{ab}Y^{ab*}_{U}Y^{ab}_{U}={\rm Tr}[Y_{U}Y_{U}^{\dagger}]$
times the trace over the gauge contractions ${\rm Tr}[\Delta\Delta^{\dagger}]$
where $\Delta\equiv{\varepsilon}^{kl}{\delta_{3}}_{j}^{i}$.
For the SM gauge group (with the properly normalized hypercharge) one then obtains
$y_{1U}=\frac{17}{10}$,
$y_{2U}=\frac{3}{2}$
and
$y_{3U}=2$,
that coincide with the values given in the first column of the matrix (B.5) in Ref.~\cite{Machacek:1983-85}.

All of the $y_{pk}$ coefficients as well as the structures of the
relevant $\Delta$-tensors are reported
in Table \ref{tab:Yukawas}.

\renewcommand{\arraystretch}{1.2}
\begin{table*}[h]
\centering
\begin{tabular}{ccccccc}
\hline
{$G_p$}  &  $y_{pk}$  &  {$k$} &  {\rm Gauge structure} &  {\rm Higgs representation} &  Tensor $\Delta$  & Tr$[\Delta\Delta^{\dagger}]$
\\
\hline
\\
$\begin{array}{c}
1_Y \\
2_L \\
3_c
\end{array}$
&
$\left(
\begin{array}{lcc}
 \frac{17}{10} & \frac{1}{2} & \frac{3}{2} \\
 \frac{3}{2} & \frac{3}{2} & \frac{1}{2} \\
 2 & 2 & 0
\end{array}
\right)$  &
$\begin{array}{c}
{\rm U} \\
{\rm D} \\
{\rm E}
\end{array}$
&
$\begin{array}{c}
\overline{Q_{L}}_{kj}U_{R}^{i}\tilde h_{l} \\
\overline{Q_{L}}_{kj}D_{R}^{i}h^{l} \\
\overline{L_{L}}_{k}E_{R}^{i}h^{l} \\
\end{array}$
&
$h^{l}: (+\frac{1}{2},2,1)$
&
$\begin{array}{c}
\epsilon^{kl}{\delta_3}_{i}^{j} \\
{\delta_{2}}^{k}_{l}{\delta_3}_{i}^{j}  \\
{\delta_{2}}^{k}_{l} \\
\end{array}$
&
$\begin{array}{c}
6 \\
6 \\
2 \\
\end{array}$
\\
\\
$\begin{array}{c}
2_L\\[0.3ex]
1_{RR} \\[0.3ex]
1_{RX}\\[0.3ex]
1_{XR}\\[0.3ex]
1_{XX}\\[0.3ex]
3_{c}
\end{array}$ &
$\left(
\begin{array}{cccc}
 \frac{3}{2} & \frac{3}{2} & \frac{1}{2} & \frac{1}{2} \\
 \frac{3}{2} & \frac{3}{2} & \frac{1}{2} & \frac{1}{2} \\
 \frac{1}{2}\sqrt{\frac{3}{2}} & -\frac{1}{2}\sqrt{\frac{3}{2}} & -\frac{1}{2}\sqrt{\frac{3}{2}} & \frac{1}{2}\sqrt{\frac{3}{2}} \\
 \frac{1}{2}\sqrt{\frac{3}{2}} & -\frac{1}{2}\sqrt{\frac{3}{2}} & -\frac{1}{2}\sqrt{\frac{3}{2}} & \frac{1}{2}\sqrt{\frac{3}{2}} \\
 \frac{1}{2} & \frac{1}{2} & \frac{3}{2} & \frac{3}{2} \\
2 & 2 & 0 & 0
\end{array}
\right)$  &
$\begin{array}{c}
{\rm U} \\
{\rm D} \\
{\rm N} \\
{\rm E}
\end{array}$ &
$\begin{array}{c}
\overline{Q_{L}}_{kj}U_{R}^{i}\tilde h_{l} \\
\overline{Q_{L}}_{kj}D_{R}^{i}h^{l} \\
\overline{L_{L}}_{k}N_{R}\tilde h_{l} \\
\overline{L_{L}}_{k}E_{R}h^{l} \\
\end{array}$
&
$h^{l}:(2,+\frac{1}{2},0,1)$
&
$\begin{array}{c}
\epsilon^{kl}{\delta_3}_{i}^{j} \\
{\delta_{2}}^{k}_{l}{\delta_3}_{i}^{j}  \\
\epsilon^{kl}\\
{\delta_{2}}^{k}_{l} \\
\end{array}$
&
$\begin{array}{c}
6 \\
6 \\
2 \\
2 \\
\end{array}$
\\
\\
$\begin{array}{c}
2_L\\
2_{R} \\
1_{X}\\
3_{c}
\end{array}$
&
$\left(
\begin{array}{ccc}
 3 & 1 \\
 3 & 1 \\
 1 & 3 \\
 4 & 0
\end{array}
\right)$  &
$\begin{array}{c}
{\rm Q} \\
{\rm L}
\end{array}$ &
$\begin{array}{c}
Q_{L}^{ik}Q^{c\,m}_{Lj}\phi^{ln} \\
L_{L}^{k}L^{c\,m}_{L}\phi^{ln} \\
\end{array}$
&
$\phi^{ln}:(2,2,0,1)$
&
$\begin{array}{c}
\epsilon_{kl}\epsilon_{mn}{\delta_3}_{i}^{j} \\
\epsilon_{kl}\epsilon_{mn} \\
\end{array}$
&
$\begin{array}{c}
12 \\
4 \\
\end{array}$
\\
\\
$\begin{array}{c}
2_L\\
1_{X}\\
4_{C}
\end{array}$
&
$\left(
\begin{array}{cc}
 2 & 2 \\
 2 & 2 \\
 2 & 2
\end{array}
\right)$  &
$\begin{array}{c}
{\rm F^{U}} \\
{\rm F^{D}}
\end{array}$
&
$\begin{array}{c}
\overline{F_{L}}_{kj}F^{Ui}_{R}\tilde h_{l} \\
\overline{F_{L}}_{kj}F^{Di}_{R}h^{l}
\end{array}$
&
$h^{l}:(2,+\frac{1}{2},1)$
&
$\begin{array}{c}
\epsilon^{kl}{\delta_4}_{i}^{j} \\
{\delta_{2}}^{k}_{l}{\delta_4}_{i}^{j}  \\
\end{array}$
&
$\begin{array}{c}
8 \\
8 \\
\end{array}$
\\
\\
$\begin{array}{c}
2_L\\
2_{R}\\
4_{C}
\end{array}$
&
$\left(
\begin{array}{c}
 4 \\
 4 \\
 4
\end{array}
\right)$  &
{\rm F} &
$F_{L}^{ik}F^{c\,m}_{Lj}\phi^{ln}$
&
$\phi^{ln}:(2,2,1)$
&
$\begin{array}{c}
\epsilon_{kl}\epsilon_{mn}{\delta_4}_{i}^{j} \\
\end{array}$
&
$\begin{array}{c}
16 \\
\end{array}$
\\
\\
\\
$\begin{array}{c}
2_L\\
1_{X}\\
4_{C}
\end{array}$
&
$\left(
\begin{array}{cc}
 \frac{15}{4} & \frac{15}{4} \\
\frac{15}{4} & \frac{15}{4} \\
\frac{15}{4} & \frac{15}{4}
\end{array}
\right)$  &
$\begin{array}{c}
{\rm F^{U}} \\
{\rm F^{D}}
\end{array}$ &
$\begin{array}{c}
\overline{F_{L}}_{kj}F^{Ui}_{R}\tilde H^{a}_{l} \\
\overline{F_{L}}_{kj}F^{Di}_{R}H^{la}
\end{array}$
 &
$H^{la}:(2,+\frac{1}{2},15)$
&
$\begin{array}{c}
\epsilon^{kl}(T_{a})_{i}^{j}\\
\delta^{k}_{l}(T_{a})_{i}^{j}\\
\end{array}$
&
$\begin{array}{c}
15 \\
15 \\
\end{array}$
\\
\\
$\begin{array}{c}
2_L\\
2_{R}\\
4_{C}
\end{array}$
&
$\left(
\begin{array}{c}
\frac{15}{2} \\
\frac{15}{2} \\
\frac{15}{2}
\end{array}
\right)$  &
{\rm F} &
$F_{L}^{ik}F^{c\,m}_{Lj}\Phi^{lna}$
&
$\Phi^{lna}:(2,2,15)$
&
$\epsilon_{kl}\epsilon_{mn}(T_{a})_{i}^{j}$
&
$\begin{array}{c}
30 \\
\end{array}$
\\
\\
\hline
\end{tabular}
\caption{The two-loop Yukawa contributions to the gauge sector $\beta$-functions
in \eq{Ycpk} are detailed. The index $p$ in $y_{pk}$ labels the gauge groups while
$k$ refers to flavour.
In addition to the Higgs bi-doublet from the 10-dimensional representation
(whose components are denoted according to the relevant gauge symmetry by $h$ and $\phi$)
extra bi-doublet components in $\overline{126}_{H}$ (denoted by $H$ and $\Phi$)
survives from unification
down to the Pati-Salam breaking scale as required by a realistic SM fermionic spectrum.
The $T_{a}$ factors are the generators of $SU(4)_{C}$ in the standard normalization.
As a consequence of minimal fine tuning, only one linear combination of $10_H$
and $\overline{126}_H$ doublets survives below the $SU(4)_{C}$ scale.
The $U(1)_{R,X}$ mixing in the case $2_L1_R1_X3_c$ is explicitly displayed.}
\label{tab:Yukawas}
\end{table*}

\end{document}